\documentclass[useAMS,usenatbib,usegraphicx]{mn2e}
\usepackage{amsmath,amsfonts,amssymb}
\usepackage[usenames,dvipsnames]{xcolor}

\def\Msun{\hbox{$\rm\, M_{\odot}$}}

\title [Simulations of the galaxy population: implications for
galactic winds and the fate of their ejecta] {Simulations of the
  galaxy population constrained by observations from $z=3$ to the
  present day: implications for galactic winds and the fate of their
  ejecta}

\author[Henriques et al.]  
{Bruno M. B. Henriques$^{1}$\thanks{E-mail:bhenriques@mpa-garching.mpg.de},
Simon D. M. White$^{1}$, Peter A. Thomas$^{2}$, 
\newauthor
Raul E. Angulo$^{3}$, Qi Guo$^{4}$, Gerard Lemson$^{1}$, Volker
Springel$^{5,6}$ \vspace{0.4cm}\\
  {}$^{1}$Max-Planck-Institut f\"ur Astrophysik, Karl-Schwarzschild-Str. 1, 85741 Garching b. M\"unchen, Germany\\
  {}$^{2}$Astronomy Centre, University of Sussex, Falmer, Brighton BN1 9QH, United Kingdom\\
 {}$^{3}$Kavli Institute for Particle Astrophysics and Cosmology, Stanford University, Menlo Park, CA 94025, USA\\
 {}$^{4}$Partner Group of the Max-Planck-Institut f\"ur Astrophysik, National Astronomical Observatories, Chinese Academy of Sciences, \\
             ~Beijing, 100012, China\\
{}$^{5}$Heidelberger Institut  f\"ur Theoretische Studien, Schloss-Wolfsbrunnenweg 35, 69118 Heidelberg, Germany\\
{}$^{6}$Zentrum f\"ur Astronomie der Universit\"at Heidelberg, ARI, M\"onchhofstr. 12-14, 69120 Heidelberg, Germany\\
}
 
\begin{document}

\date{Submitted to MNRAS}

\pagerange{\pageref{firstpage}--\pageref{lastpage}} \pubyear{2012}

\maketitle

\label{firstpage}

\begin{abstract}
  We apply Monte Carlo Markov Chain (MCMC) methods to large-scale
  simulations of galaxy formation in a $\Lambda$CDM cosmology in order
  to explore how star formation and feedback are constrained by the
  observed luminosity and stellar mass functions of galaxies.  We
  build models jointly on the Millennium and Millennium-II
  simulations, applying fast sampling techniques which allow observed
  galaxy abundances over the ranges $7<\log M_\star/\Msun < 12$ and
  $0\leq z\leq 3$ to be used simultaneously as constraints in the MCMC
  analysis. When $z=0$ constraints alone are imposed, we reproduce the
  results of previous modelling by \citet{Guo2012}, but no single set
  of parameters can reproduce observed galaxy abundances at all
  redshifts simultaneously, reflecting the fact that low-mass galaxies
  form too early and thus are overabundant at high redshift in this
  model. The data require the efficiency with which galactic wind
  ejecta are reaccreted to vary with redshift and halo mass quite
  differently than previously assumed, but in a similar way as in some
  recent hydrodynamic simulations of galaxy formation.  We propose a
  specific model in which reincorporation timescales vary inversely
  with halo mass and are independent of redshift. This produces an
  evolving galaxy population which fits observed abundances as a
  function of stellar mass, $B$- and $K$-band luminosity at all
  redshifts simultaneously. It also produces a significant improvement
  in two other areas where previous models were deficient. It leads to
  present day dwarf galaxy populations which are younger, bluer, more
  strongly star-forming and more weakly clustered on small scales than
  before, although the passive fraction of faint dwarfs remains too
  high.

\end{abstract}


\section{Introduction}
\label{sec:intro}

The semi-analytic modelling technique uses a set of parametrised,
physically based equations to describe each of the astrophysical
processes that affect galaxy formation and evolution \citep{White1989,
  Cole1991, Lacey1991, White1991}. The development of the galaxy
population as a whole is followed by applying these recipes to a set
of merger trees representing the hierarchical assembly of nonlinear
structure in the dark matter distribution. Such halo merger trees can
be constructed efficiently using extensions of the \cite{Press1974}
formalism \citep[e.g.][]{Cole1991, Kauffmann1993, Cole1994,
  Somerville1999} but if spatial and kinematic information is needed,
for example, for analysis of galaxy clustering, it is more effective
to extract them directly from numerical simulations
\citep{Roukema1997, Kauffmann1999, Helly2003, Hatton2003}. With increasing
numerical resolution it becomes possible to base the trees not on the
haloes themselves but rather on the individual gravitationally
self-bound subhaloes from which they are built \citep{Springel2001,
  Springel2005, Kang2005, DeLucia2006}. This increases the fidelity
with which the galaxy distribution is simulated, but introduces a
dependence on numerical resolution which must be tested through
careful convergence studies \citep[e.g.][]{Springel2001, Guo2011}.

The great advantage of such semi-analytic simulations is their
decoupling of the computation of dark matter evolution from that of
the baryonic components.  The physically consistent but relatively
simple models adopted for complex and poorly understood baryonic
processes allow a wide range of possible descriptions to be explored.
The optimal parameters for each can be determined in an acceptable
amount of computer time, allowing the different modelling assumptions
to be tested by detailed comparison with the relevant observational
data. At present, semi-analytic methods are the only technique able to
simulate the evolution of the galaxy population on a scale and with a
precision which allows detailed interpretation of modern surveys of
the galaxy population. The price, of course, is that they provide at
best only very crude information about the internal structure of
galaxies or the structure of the intergalactic medium.

The latest version of the Munich semi-analytic model
\citep[][hereafter G11]{Guo2011, Guo2012} was tested and its
parameters adjusted by comparison with low-redshift galaxy data,
namely, stellar mass and optical luminosity functions, gas fractions,
colours, gas-phase metallicities, bulge-to-disk ratios, bulge and disk
sizes, and the black hole -- bulge mass relation. Predictions were
then tested by looking at satellite abundances, clustering, the
evolution of the stellar mass function and the integrated star
formation history. This analysis revealed a good match between model
and data, particularly for the present-day universe, but some
discrepancies remained. Most notable were an excess of lower mass
galaxies at high redshift \citep[also visible in the $K$-band
luminosity function of][]{Henriques2012}, a substantial population of
red dwarfs at low redshift, where very few such galaxies are observed,
and excessive small-scale clustering of low-redshift dwarf
galaxies. \citet{Guo2012} showed that only the last of these problems
is significantly alleviated if a WMAP7 cosmology is adopted instead of
the WMAP1 cosmology of the original simulations. Similar problems are
found in most recent semi-analytic galaxy formation models
\citep{Fontanot2009, Cirasuolo2010, Henriques2011, Somerville2011,
  Bower2012, Weinmann2012}, and can thus be viewed as generic.

In this paper we revisit the predictions of the G11 semi-analytic
model using the Monte Carlo Markov Chain (MCMC) methods introduced by
\citet{Henriques2009} and \citet{Henriques2010} (see
\citealt{Kampakoglou2008, Bower2010, Lu2010} for related methods). We
constrain parameters using observational estimates of the stellar mass
function and of the $K$- and $B$-band luminosity functions at a
variety of redshifts.  The MCMC approach offers a systematic and
objective means to identify the regions of parameter space consistent
with the data and to obtain robust best-fit parameter estimates with
associated confidence ranges.  This can yield new insight into how the
data constrain galaxy formation physics.

We start by investigating the uniqueness of the parameter choices of
G11, finding that they are indeed nearly optimal when representing 
low-redshift data. We then use the MCMC technique to identify the
parameter regions preferred by the observational data at other
redshifts, testing whether they overlap with each other and with the
region preferred by the low-redshift data. Although almost all the G11
parameters fall in the allowed region at all redshifts, the rate at
which gas is reincorporated after ejection in a wind is required to
change.  Re-accretion must be less efficient than originally assumed
at early times, and more efficient at late times. We use the MCMC
scheme to identify a simple modification of the treatment of this
process which can explain the data simultaneously at all redshifts. We
then test this prescription by comparing its predictions with a broad
set of observations including stellar metallicities, galaxy colours,
star formation rates and ages, clustering and the evolution of the
stellar mass-halo mass relation.

This work uses merger trees from two very large dark matter
simulations, which are well adapted to study galaxy formation physics
on a variety of scales. The Millennium Simulation \citep{Springel2005}
follows $10^{10}$ particles in a cube of side $500h^{-1}\rm{Mpc}$,
implying a particle mass of $8.6\times10^{8}h^{-1}\Msun$, while the
Millennium-II Simulation \citep{Boylan2009} uses the same number of
particles in a region a fifth the linear size, resulting in 125 times
better mass resolution.  Combined, the two simulations produce a
galaxy formation model with a dynamic range of five orders of
magnitude in stellar mass. The distribution of physical properties
converges for galaxies with $10^{9.5}\Msun < \rm{M}_{\star} <
10^{11.5}\Msun$.  The simulations can be scaled from the original
WMAP1 cosmology to the currently preferred WMAP7 cosmology using the
technique of \citet{Angulo2010}. Galaxy populations almost identical
to those in the original G11 model can then be obtained through small
adjustments to the semi-analytic modelling parameters \citep{Guo2012}.
This scaling changes the box-size and particle mass of the
simulations, as well as shifting their time axis \citep[see][for
  details]{Guo2012}. Below we work in the WMAP7 cosmology using the
scaled merger trees.

The physical models used here for processes such as cooling, star
formation and feedback have been developed gradually within the
``Munich'' galaxy formation model over many years \citep{White1991,
  Kauffmann1993, Kauffmann1999, Springel2001, Springel2005}. The
current G11 version includes a robust treatment of supernova feedback
(following \citealt{Delucia2004b} rather than \citealt{Croton2006},
see also \citealt{Benson2003}). It follows the growth of black holes
through accretion and merging \citep{Kauffmann2000} and the quenching
of star-formation by AGN feedback \citep{Croton2006}.  It treats
separately dust extinction from the inter-stellar medium and from
young birth clouds \citep{DeLucia2007} and it can predict luminosities
over a wide wavelength range using two different stellar population
synthesis models \citep[see][]{Henriques2011, Henriques2012}. G11
changed several of the baryonic physics recipes in order to improve
the treatment of dwarf and satellite galaxies.  Supernova feedback was
increased in low-mass galaxies, and environmental effects on
satellites were treated more realistically.  They also improved the
tracking of angular momentum as material moves between the various
gaseous and stellar components, allowing a better treatment of the
sizes of disks and bulges.

This paper is organized as follows. In Section~\ref{sec:mcmc}, we
describe our implementation of the MCMC method.
Section~\ref{sec:previous_model} applies this method to the G11 model,
constraining parameters independently with data at each of a set of
redshifts. No parameter set is consistent with the data at all
redshifts. Guided by these MCMC results, we formulate in
Section~\ref{sec:new_model} a modification of the G11 model which can
fit all redshifts simultaneously. The predictions of this new model
are compared to the constraining observations in
Section~\ref{sec:results}, while Section~\ref{sec:add_observations}
tests it against observations of additional properties that were not
used as constraints. In Section~\ref{sec:conclusions}, we summarize
our conclusions. Appendix~\ref{app:physics} sets out the equations
that define the parameters explored using MCMC sampling, while
Appendix~\ref{app:sampling} describes how we construct a
representative sample of dark matter trees which allows us to obtain
results rapidly for any specific semi-analytic model.
Appendix~\ref{app:obs_err} gives details of the observational data we
use and of the uncertainties we assume for them, when using MCMC
techniques to constrain model parameters.

A recent preprint by \citet{Mutch2012} used MCMC techniques to explore
similar issues using the model of \citet{Croton2006} applied to the
Millennium Simulation in its original WMAP1 cosmology. Because they
used stellar mass functions alone, and did not consider epochs earlier
than $z=0.83$, the evolutionary problem addressed here was only weakly
present in their observational constraints. As a result they came to
somewhat different conclusions than we do in this paper.

Throughout this work we use the \citet{Maraston2005} stellar
population synthesis model, but we have checked that for all the
properties considered here the \citet{Charlot2007} code gives very
similar results. As noted above, we assume a WMAP7 cosmology with
specific parameters $h=0.704$, $\Omega_m=0.272$,
$\Omega_{\Lambda}=0.728$, $n=0.961$, and $\sigma_8=0.81$, using the MS
and MS-II merger trees scaled exactly as in \cite{Guo2012}.

\section{Monte Carlo Markov Chains}
\label{sec:mcmc}
Semi-analytic models simulate a large number of complex physical
processes whose interplay shapes the distribution of galaxy
properties.  As a result, it is not straightforward to find the model
that best fits a given set of observations, and to understand how
unique a specific prescription or parameter set is. This becomes even
more problematic when treatments are modified to allow interpretation
of new kinds of observation, because such modifications often
destroy the match to existing data. In addition, when reasonable
agreement appears impossible to achieve, it is hard to know whether
this reflects a failure to find the correct parameter set in a
high-dimensional space, or overly simplistic modelling of one or more
of the processes treated, or the omission of an important process, or
an issue with the parent dark matter simulation or the structure
formation paradigm underlying it.

Many of these problems can be mitigated by combining constraints from
multiple observations of a wide range of galaxy properties with proper
sampling of the high-dimensional model parameter space. Monte Carlo
Markov Chain (MCMC) algorithms can be used to clarify how individual
parameters influence specific galaxy properties, to obtain confidence
limits for parameters, and to quantify the agreement between model and
observation in a statistically robust way. They were first applied to
the Munich galaxy formation model by \citet{Henriques2009} and
\cite{Henriques2010}. Such algorithms sample the allowed parameter
regions at a rate proportional to the posterior probability of the
model conditioned by the observational constraints, making them very
efficient in analysing high-dimensional spaces.  Section~3 of
\citet{Henriques2009} gives a full description of the
Metropolis-Hastings algorithm used here \citep{Metropolis1953,
  Hastings1970} which builds chains in which each step to a new set of
parameters depends only on the current parameter set, not on the
previous history. Appendix~\ref{app:sampling} gives some technical
details of our implementation. Of particular relevance is our
identification of a subset of merger trees which allows rapid
convergence to results which are statistically representative of the
full cosmological volume. With this approach, we find $\sim1\%$ of the
full data to be sufficient to reproduce the stellar mass function of
the combined Millennium and Millennium-II simulations to the accuracy
we need, speeding up the MCMC calculations by about a factor of 100
and allowing us to constrain the model using the observed abundance of
galaxies with stellar masses ranging from $10^7\Msun$ to
$10^{12}\Msun$.


\subsection{Observational constraints}
\label{subsec:obs_constraints}

A crucial aspect of semi-analytic modelling, and of MCMC analysis in
particular, is the choice and characterisation of the observational
data used as constraints. The allowed regions in parameter space and
the ability to find a good simultaneous fit to multiple data sets
depend sensitively on the error bars assigned to the observations.
Purely statistical errors are well defined and accurately stated in
many observational studies, but it is much harder to account
appropriately for residual systematic uncertainties. Unrecognised
systematics can lead to apparent inconsistencies between different
determinations of the same population property, jeopardising a
meaningful comparison with theoretical predictions.  

Different approaches to this problem have been adopted in earlier
work.  For example, \citet{Bower2010} tested their model against a
single data set to which they assigned a confidence region reflecting
their own assessment of its systematic uncertainty.
\citet{Henriques2009} and \citet{Henriques2010} use multiple ``good''
determinations of each observational property, taking the scatter
among them (together with the quoted statistical errors) to indicate
likely systematic uncertainties. Here we adopt a version of the latter
approach, noting that it still involves arbitrary judgements based on
incomplete understanding, so that formal levels of agreement or
disagreement between theory and observation should be treated with
caution. In section \ref{sec:results}, theoretical models are compared
with combined data sets where we have used the quoted purely
statistical errors and the scatter among different ``good''
measurements to make a subjective assessment of the effective
uncertainties.  In Appendix \ref{app:obs_err} we show the individual
data sets and provide further details on how they were combined to
provide constraints for each property at each redshift.

The allowed region of parameter space for a given model depends not
only on the uncertainties assigned to each observation, but also, of
course, on the particular observational quantities taken as
constraints. This choice should recognise both the discriminative
power and the robustness of the data. Use of complementary types of
data can increase the aspects of the model which are significantly
constrained. In the analysis of this paper we use the stellar mass
function, and the $K$- and $B$-band luminosity functions at redshifts
$3$, $2$, $1$ and $0$. The abundance of galaxies as a function of
stellar mass is one of the most fundamental properties of the
population, but it has the disadvantage that it is not directly
observable, requiring assumptions about stellar populations and
reddening. The $i$-band luminosity function would be a useful
surrogate, since it is insensitive to dust and is unaffected by
emission from stars on the poorly understood Thermally Pulsing -
Asymptotic Giant Branch (TP-AGB), but few observational studies are
available at redshifts other than zero, making it hard to assess the
systematic uncertainties. We therefore use determinations of the
stellar mass function itself. When needed, we follow
\cite{Sanchez2011} in applying a correction of $\Delta M_\ast =-0.14$
to go from \cite{Bruzual2003} to \cite{Maraston2005} stellar
populations at $z\geq1$. As previously mentioned, the combined
dynamical range of the Millennium and Millennium-II simulations allows
us to compare model and data directly, even for the lowest stellar
masses observed at $z=0$.

A critical test of galaxy formation models comes from comparison with
the observed, redshift-dependent distributions of colour and star
formation rate. We do not include direct observational estimates of
these properties in our MCMC sampling, however, since colours are very
sensitive to the details of dust modelling and observational estimates
of star formation rates have large and redshift-dependent systematic
uncertainties which are difficult to characterize due to the small
number of observational studies available.  Instead, we include colour
information indirectly by supplementing our stellar mass function data
with estimates of rest-frame $B$- and $K$-band luminosity
functions. This provides some leverage on the star formation history,
dust content, metallicity and TP-AGB emission of galaxies.

\begin{figure*}
\centering
\includegraphics[scale=0.55, trim=0 80 0 20]{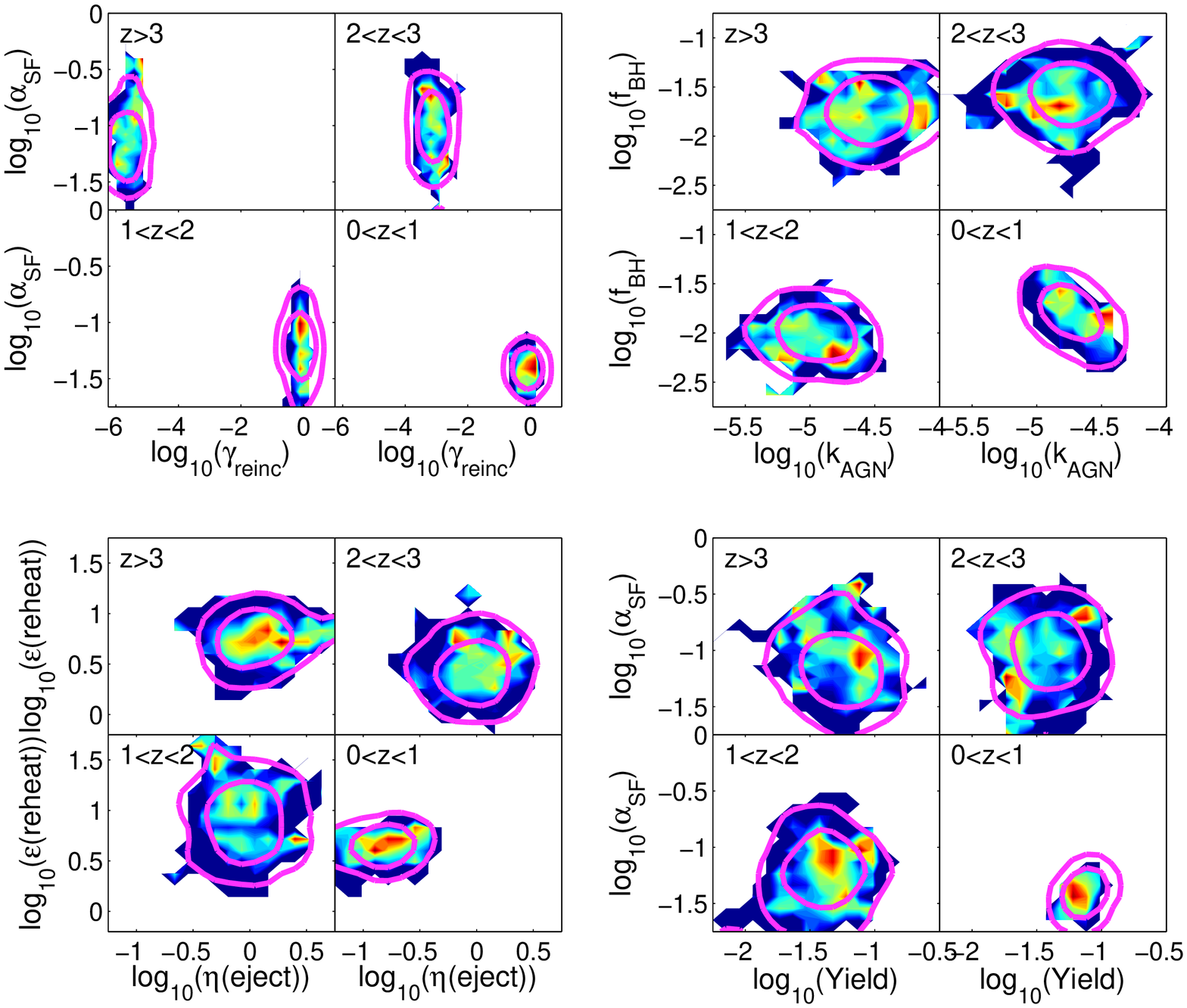}
\includegraphics[width=14.0cm, trim=0 350 0 0]{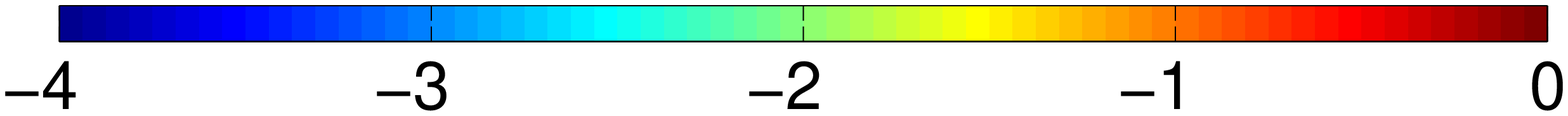}
\caption{Allowed likelihood regions for various semi-analytic
  parameter pairs at four different redshifts. Star formation
  efficiency ($\alpha_{\rm{SF}}$) and gas reincorporation efficiency
  ($\gamma$) are shown in the four upper left panels; black
  hole growth efficiency ($f_{\rm{BH}}$) and AGN radio mode efficiency
  ($k_{\rm{AGN}}$) in the four upper right panels; gas reheating
  efficiency ($\epsilon$) and gas ejection efficiency ($\eta$) in the
  four lower left panels; star formation efficiency
  ($\alpha_{\rm{SF}}$) and metal yield ($z_{\rm{yield}}$) in the four
  lower right panels. The \citet{Guo2011,Guo2012} model is constrained
  independently at each of the four redshifts using observational
  estimates of the stellar mass function and the $K$- and $B$-band
  luminosity functions. Each (logarithmic) parameter space is divided
  into equal-area pixels and 2D histograms made from our MCMC
  sampling. Solid magenta lines then show equal likelihood contours
  containing 68\% and 95\% of all samples (e.g. of the marginalised 2D
  posterior distribution). Colours indicate the maximum likelihood
  value in each bin (the profile distribution) normalized by the
  global maximum likelihood. The colour scale is labelled by
  $\log_{10} L/L_{max}$.}
\label{fig:mcmc}
\end{figure*}

For each property, we combine determinations based on wide and
narrow surveys in order to obtain constraints over a large dynamic
range. The constraints used for the stellar mass, $K$-band and
$B$-band luminosity functions at different redshifts are shown in
Figures \ref{fig:smf}, \ref{fig:kband} and \ref{fig:bband},
respectively. Appendix~\ref{app:obs_err} explains how these were
obtained from various observational studies. The likelihood of the model 
given the adopted observational constraints on each individual property
is defined as:
\begin{equation} \label{eq:chi_p}
  \mathcal{L} \propto \exp -\chi^2/2,
\end{equation}
where $\chi^2$ is given by:
\begin{equation} \label{eq:chi}
  \chi^2=\sum_{i=0}^{N_{\phi}}\left(\frac{\phi_{\rm{model}}(i)-\phi_{\rm{obs}}(i)}{\sigma_{\rm{obs}}(i)}\right)^2,
\end{equation}
and $\phi_{\rm{model}}$ and $\phi_{\rm{obs}}$ are respectively the
logarithms of the number densities of galaxies in each bin as
predicted by the model and as determined observationally. 
$\sigma_{\rm{obs}}(i)$ is our characterisation of the
observational uncertainty in $\phi_{\rm{obs}}(i)$ (see
Appendix~\ref{app:obs_err}).  The final likelihood we assign to the
model is then the product of the likelihoods for each galaxy property,
either at a single redshift, or combining all redshifts:
\begin{equation} \label{eq:like}
  \mathcal{L}=\prod_{z}\left(\mathcal{L}[\rm{SMF}(z)]\mathcal{L}[{\rm LF}_K(z)]\mathcal{L}[{\rm LF}_B(z)]\right) \propto  \exp -\chi_t^2/2,
\end{equation}
where $\chi_t^2$ is the sum of the $\chi^2$ values for all the
individual stellar mass and luminosity functions\footnote{Note that we
  neglect correlations in the scatter between different data points
  and different data sets, even though these undoubtedly exist and may
  be important.}.  For a single redshift, this approach is a formal
version of that used, for example, by G11. When all redshifts are
combined, the galaxy formation physics are constrained in a
statistically robust way which takes full advantage of the fact that
the semi-analytic model links galaxy populations self-consistently
across cosmic time using a deterministic and physically based
treatment of the evolution of individual galaxies.

\subsection{Parameters}
\label{subsec:parameters}

We allow a total of 11 parameters to vary when sampling the parameter
space of the G11 model. These are: the star formation efficiency
($\alpha_{\rm{SF}}$); the black hole growth efficiency
($f_{\rm{BH}}$); the AGN radio mode efficiency ($k_{\rm{AGN}}$); three
parameters governing the reheating and injection of cold disk gas into
the hot halo phase by supernovae, the gas reheating efficiency
($\epsilon$), the reheating cutoff velocity ($V_{\rm{reheat}}$) and
the slope of the reheating dependence on $V_{\rm{vir}}$ ($\beta_{1}$);
three parameters governing the ejection of hot halo gas to an external
reservoir, the gas ejection efficiency ($\eta$), the ejection cutoff
velocity ($V_{\rm{eject}}$) and the slope of the ejection dependence
on $V_{\rm{vir}}$ ($\beta_{\rm{2}}$); the efficiency of
reincorporating gas from the external reservoir to the hot halo
($\gamma$) or the gas return time in our new reincorporation
model ($\gamma'$); and the yield of metals returned to the
gas phase by stars ($z_{\rm{yield}}$). The equations describing these
processes and defining the parameters are described concisely in
Appendix~\ref{app:physics}. The very fact that such a high-dimensional
parameter space can be analysed at all in a reasonable amount of
computer time emphasizes a key strength of the MCMC method. Using
naive techniques would require computing $\sim 10^{11}$ models to
survey the space with just 10 points along each parameter
dimension. In contrast, we achieve convergence for chains with as
few as 30,000 steps, although we always run at least 100,000 steps and
discard the first 10\% as a ``burn in'' phase.

\section{Testing the G11 model}
\label{sec:previous_model}

We begin by using our MCMC method to explore the regions of parameter
space for which the G11 galaxy formation model, applied within a WMAP7
cosmology as in \citet{Guo2012}, is consistent with the observational
constraints at each redshift, {\em independent} of the constraints at
the other redshifts.  It is interesting that in each case the
high-likelihood region is quite localised in the full 11-dimensional
space. The observational data at a single redshift are already
sufficient to constrain all the parameters without major degeneracies.
Furthermore, at $z=0$ the high-likelihood region includes the
parameter set chosen in the earlier paper, showing that the less
formal tuning procedure adopted there did, in fact, result in
near-optimal parameters.  Comparing preferred parameter space regions
at different redshifts we find very substantial overlap. Only one
of the 11 parameters, the efficiency $\gamma$ with which
ejected gas from the external reservoir is reincorporated into the hot
halo, is required to take substantially different values at different
times.

This can be seen clearly in Fig.~\ref{fig:mcmc} which shows the
posterior likelihood distributions at the four redshifts marginalised
into four different two-parameter subspaces: star formation efficiency
and gas reincorporation efficiency in the upper left panels; black
hole growth efficiency and AGN radio mode efficiency in the upper
right panels; gas reheating efficiency and gas ejection efficiency in
the lower left panels; and star formation efficiency and metal yield
in the lower right panels. In each panel, equal likelihood contours
containing 68\% and 95\% of the marginalised posterior distribution
are shown as thick magenta lines, while colours indicate the maximum
likelihood over all the MCMC samples in each pixel. In fact, the 95\%
regions for seven of the eight parameters overlap at least marginally
across all four redshifts, whereas the reincorporation efficiency
($\gamma$) is required to evolve strongly. The observations indicate
that very little ejected gas is reincorporated at high redshift, but
that reincorporation is substantially more effective at $z\leq
2$. This appears required within the G11 representation of galaxy
formation physics in order to obtain the observed low abundances of
lower mass galaxies at $z\geq 2$ while maintaining the higher
abundances observed for similar mass galaxies at lower redshift. There
is some indication that the supernova ejection efficiency $\eta$
should be lower at $z\sim 0$ than at higher redshift (x-axis on the
bottom left panels) but the variation required is much smaller than
the 4 orders of magnitude needed for $\gamma$ and a reasonable
compromise value can be found on the edge of the 95\% regions ($\log
\eta \sim -0.3$).

It is worth stressing the discerning power of this procedure. It
indicates that most of the physical assumptions made in the G11 model
are consistent with the observed evolution of the galaxy population
over $3>z>0$. In addition, it clearly indicates which of the
assumptions need to be modified and how they should be corrected. The
apparent lack of degeneracies also suggests that, for most of the
efficiency parameters, evolutionary trends other than those currently
assumed are disfavored by the observational constraints we have
adopted. For example, a star formation efficiency that increases with
decreasing redshift, as proposed by \citet{Wang2012} to match the
evolution of the number density of dwarfs, is not consistent with the
data in the context of \emph{our} model. Indeed, in this model star
formation efficiency primarily affects the gas-to-star ratios of
dwarfs rather than their stellar masses or star formation rates.  As
in some hydrodynamical simulations \citep{Oppenheimer2008, Haas2012},
gas simply accumulates in a dwarf until it is sufficient to fuel star
formation and ejection rates which balance the gas supply through
infall. Similarly, changing the scaling of wind properties
  alone cannot produced the required evolution, as recently found also
  by \citet{Bower2012}.

\section{A new model for reincorporation of ejected gas}
\label{sec:new_model}

The G11 version of the Munich semi-analytic model assumed the gas
ejected from haloes to be stored in an external reservoir and to be
reincorporated into the system on a timescale which depends on halo
mass and redshift. The particular model adopted was,
\begin{equation} \label{eq:reincorporation_guo1}
  \dot{M}_{\rm{ejec}}=-\frac{M_{\rm{ejec}}}{t_{\rm{reinc}}},
\end{equation}
with
\begin{equation} \label{eq:reincorporation_guo2}
  t_{\rm{reinc}}=\frac{1}{\gamma}t_{\rm{dyn,h}}\left(\frac{V_{\rm{vir}}}
    {220\, {\rm km\,s^{-1}}}\right)^{-1},
\end{equation}
where $M_{\rm{ejec}}$ is the current amount of gas in the reservoir,
$t_{\rm{dyn,h}}=0.1H(z)^{-1}$ is the dynamical time of the halo,
$V_{\rm{vir}}$ is its virial velocity and $\gamma$ is a free
parameter. The $V_{\rm{vir}}$ dependence was introduced by G11 to
model the higher wind velocities (relative to escape velocity) in
lower mass systems. The scaling with $t_{\rm{dyn,h}}$ imposes a
redshift dependence similar to that of the dark matter growth time of
the haloes, which is nearly independent of mass but decreases quite
strongly at earlier times \citep[e.g.][]{Guo2008}.  Our MCMC analysis
shows, however, that the data require $\gamma$ to be much smaller (and
thus reincorporation to be much less effective) at $z\geq 2$ than at
lower redshifts.  In order to find a description of the
reincorporation process that can reproduce the data at all times, we
re-write equation~(\ref{eq:reincorporation_guo2}) as:
\begin{equation} \label{eq:reincorporation_z2}
 t_{\rm{reinc}}=\frac{1}{\gamma}(1+z)^{\delta_1}V_{\rm{vir}}^{-\delta_2}~~t_{\rm{dyn,h}},
\end{equation}
and we re-run our MCMC chains constrained simultaneously by the
observations at all redshifts and including $\delta_1$ and $\delta_2$
as free parameters.

\begin{figure}
\centering
\includegraphics[scale=0.2]{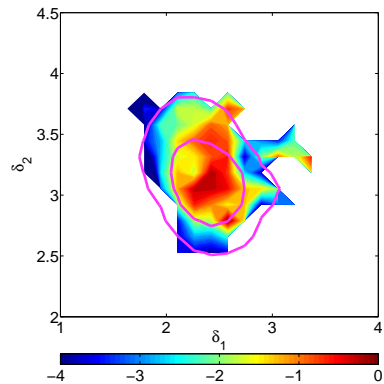}
\caption{High likelihood region for the $\delta_1$ and $\delta_2$
  parameters in the reincorporation model described by equation
  (\ref{eq:reincorporation_z2}). The model is constrained using the
  observed stellar mass functions and the $K$- and $B$-band luminosity
  functions at $z=0$, 1, 2 and 3 simultaneously. The solid lines are
  equal likelihood contours containing 68\% and 95\% of the
  marginalised 2D posterior distribution, while colours indicate the
  maximum likelihood value over all MCMC parameter samples projected
  within each pixel.  The logarithmic colour scale is normalized to
  the global maximum likelihood value.}
\label{fig:z_search}
\end{figure}

The results of this procedure are shown in Fig.~\ref{fig:z_search}. 
After marginalising over other parameters the high likelihood region
is quite localised in the $\delta_1$--$\delta_2$ plane. The maximum likelihood
values of these parameters and their $\pm 1\,\sigma$ range are
$\delta_1=2.40^{+0.21}_{-0.33}$ and
$\delta_2=3.07^{+0.26}_{-0.26}$. All other parameters remain in their plausible
ranges for this maximum likelihood model. If we adopt $\delta_1=2.4$ and
$\delta_2=3$  we find, using $M_{\rm{vir}}\propto V_{\rm{vir}}^3/H(z)$ 
and $t_{\rm{dyn,h}}\propto H(z)^{-1}$ that
\begin{equation} \label{eq:reinc_final_z}
t_{\rm{reinc}}\propto \frac{1}{M_{\rm{vir}}}~\frac{(1+z)^{2.4}}{\Omega_m(1+z)^3 +(1-\Omega_m)},
\end{equation}
where $\Omega_m\sim 0.25$ is the present-day matter density. The redshift-dependent
ratio on the right-hand-side of this equation varies by less than a factor of 2
for $0\leq z \leq 6$ so we neglect it and adopt a simpler reincorporation model
in which the relevant timescale 
is inversely proportional to halo mass and independent of redshift:
\begin{equation} \label{eq:reinc_final}
t_{\rm{reinc}}=-\gamma' \frac{10^{10}\Msun}{M_{\rm{vir}}},
\end{equation}
where the constant $\gamma'$ now has units of time. Using our MCMC
chains to identify the best-fit of this new model to our constraining
data at all four redshifts, we find $\chi_t^2=127$ for 134 degrees of
freedom and 11 adjusted parameters. For comparison, applying the same
procedure to the G11 model yields $\chi_t^2=475$ for the same number
of adjusted parameters. Thus our modified assumptions about
reaccretion of ejected gas substantially improve the ability of the
model to represent the observational data.

\begin{figure*}
\centering
\includegraphics[width=17.9cm]{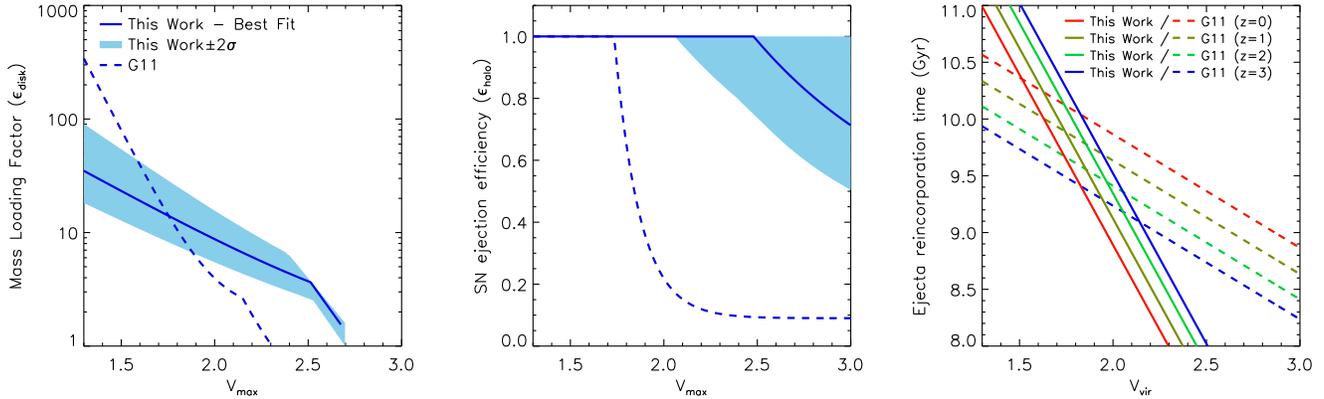}
\caption{ Illustration of the major changes in the dependence of
  feedback on galaxy properties between the G11 model and the model of
  this paper. The left panel shows the disk reheating efficiency
  $\epsilon_{\rm disk}$ as a function of maximum circular velocity
  $V_{\rm max}$. Often referred to as the mass-loading factor, this is
  the ratio of the star formation rate to the rate at which ISM
  material is heated and injected into the hot halo. The middle panel
  shows the halo ejection efficiency $\epsilon_{\rm halo}$ as a
  function of $V_{\rm max}$.  This is the fraction of the available SN
  energy which is used in reheating disk gas and in ejecting hot gas
  from the halo. The right panel shows the reincorporation timescale
  $t_{\rm reinc}$ as a function of halo virial velocity $V_{\rm vir}$
  and of redshift. In each panel dashed lines refer to the G11 model
  and solid lines to our new model with its best-fit parameters. The
  blue shaded regions in the left two panels give the $2\sigma$ range
  allowed by our MCMC sampling. Colours in the right panel indicate
  redshift as shown by the label.}
\label{fig:feedback_scalings}
\end{figure*}

The best-fit value of the reincorporation time-scale,
$\gamma'=1.8\times10^{10}\rm{yr}$, implies that haloes with
$M_{\rm{vir}}\geq 10^{11.5}\Msun$ have very short reincorporation
times, hence are able to eject very little gas at any relevant
redshift. In contrast, very little of the gas ejected from haloes with
$M_{\rm{vir}}\leq 10^{10}\Msun$ ever returns. For intermediate-mass
haloes, ejection is most effective at high redshift because at given
halo mass, cooling, star formation and feedback are all typically
stronger at earlier times.  We note that higher halo masses at higher
redshift (see Fig.~\ref{fig:mstar_mvir}) result in galaxies of given
stellar mass having gas return times which are shorter at early times
than at $z=0$. However, the redshift dependence is now much weaker
than in G11, and the reincorporation time-scales in low-mass galaxies
are longer, ensuring that gas is predominantly returned at low
redshift.  The changes are illustrated in the right panel of
Fig.~\ref{fig:feedback_scalings}; at $z=3$ the reincorporation time is
longer in the new model than in G11 for all haloes with $V_{\rm
  vir}<140$~km/s, whereas at $z=0$ this is only true for $V_{\rm
  vir}<30$~km/s.

\begin{table*}
\begin{center}
\caption{Statistics from the MCMC parameter estimation. The best-fit
  values of parameters and their ``2-$\sigma$'' confidence limits are
  compared with the values published in \citet{Guo2012} for their
  WMAP7 model constrained by $z=0$ data alone.}
\label{table:margestats}
\begin{tabular}{lccccc}
\hline
\hline
&Guo12 &New Model  &2$\sigma$ lower limit  &2$\sigma$ upper limit\\
\hline
\hline
$\alpha_{\rm{SF}}$ (SF efficiency)&0.01 &0.055 &0.027 &0.060\\
\hline
$k_{\rm{AGN}}$ (Radio feedback efficiency) &$7.0\times10^{-6}$ &$3.2\times10^{-5}$ &$2.7\times10^{-5}$ &$4.1 \times10^{-5}$\\
$f_{\rm{BH}}$ (BH growth efficiency)&0.03 &0.015 &0.011 &0.017 \\ 
\hline
$\epsilon$ (mass-loading efficiency)&4.0 &2.1 &1.8 &2.6 \\
$V_{\rm{reheat}}$ (mass-loading scale)&80 &405 &315 &473 \\
$\beta_{1}$ (mass-loading slope)&3.2 &0.92 &0.82 &1.14 \\
\hline
$\eta$ (SN heating efficiency)&0.2 &0.65 &0.52 &0.87\\
$V_{\rm{eject}}$ (SN heating scale)&90. &336 &263 &430\\
$\beta_{2}$ (SN heating slope)&3.2 &0.46 &0.38 &0.59\\
\hline
$\gamma^\prime$ (ejecta reincorporation)& not applicable  &$1.8\times10^{10}\rm{yr}$ &$1.8\times10^{10}\rm{yr}$ &$1.9\times10^{10}\rm{yr}$  \\
\hline
$Z_{\rm{yield}}$ (metals yield) &0.03 &0.047 &0.040 &0.051 \\
\hline
\hline
\end{tabular}
\end{center}
\end{table*}

Our proposed dependence of the gas return time-scale on halo mass can
also be compared with that found in recent direct numerical
simulations of the physics of feedback. \citet{Oppenheimer2008} found
that the gas ejected in winds is affected not only by gravitational
forces, but also by ram pressure against the circumgalactic medium. As
a result, material is less likely to escape in denser
environments. These hydrodynamical simulations suggest that the time
for ejected material to return to a galaxy scales inversely with the
total mass of the system rather than with the local dynamical time, a
very similar result to that we find here. More specifically,
\citet{Oppenheimer2010} found return times to scale with
$M_{\rm{vir}}^{-1.5}$ for constant wind speed and with
$M_{\rm{vir}}^{-0.6}$ for wind speeds proportional to the virial
velocity of the system. The latter is more similar to our feedback
implementation (see Appendix~\ref{app:physics} for details). Indeed,
we obtain a compatible scaling of the return times in our new
reincorporation model. Our preferred value for $\gamma'$
($1.8\times10^{10}\rm{yr}$) results in a reincorporation time that
varies from $1.8\times10^{10}\rm{yr}$ for haloes with
$M_{\rm{vir}}=10^{10}\Msun$ to $1.8\times10^8\rm{yr}$ for haloes with
$M_{\rm{vir}}=10^{12.}\Msun$. These correspond to the time-scales for
$\sim$10\% of the gas to be reincorporated in the simulation of
\citet{Oppenheimer2010}. As expected, our time-scales are shorter
since they correspond to the time for the gas to return from the
ejected to the hot phase, whereas in the hydrodynamical simulations
they correspond to the time for a full cycle, beginning when the
material leaves the star-forming region and ending when it is
reincorporated into the interstellar medium of the galaxy.

In passing we note that our new version of the semi-analytic code also
includes a number of technical changes mostly related to
book-keeping. These are designed to assure accurate mass and metal
conservation at all times, and improved memory management at run-time.
These changes can have an impact on a few individual galaxies but they
leave the global properties of the population unchanged. We have also
introduced a minor modification to the physics of infall. In previous
versions of the model, if the $M_{\rm vir}$ value of an FoF halo drops
during its evolution, then hot gas (including metals) was removed so
that the total baryon content remains equal to the cosmic baryon
fraction times $M_{\rm vir}$.\footnote{The $M_{\rm vir}$ values of FoF
  haloes generally increase with time but can drop temporarily through
  numerical fluctuations or if they pass close to a larger halo.}  We
relax this assumption here and allow each FoF halo to retain all its
associated baryons even if this puts its baryon fraction temporarily
above the cosmic value. Its baryon content then begins to increase
again only after its $M_{\rm vir}$ rises above the previous maximum
value. This change simplifies the treatment of chemical enrichment,
allowing detailed conservation of the mass of every element as
material shifts between the different baryonic components in each FoF
halo.  It also has very little impact on global galaxy properties.

\subsection{Best-fit parameters}
\label{subsec:bestfit}

As noted above, with the new reincorporation model of
equation~(\ref{eq:reinc_final}) it is possible to find parameters that
produce acceptable agreement with the observational constraints at all
the redshifts we consider. The resulting best-fit parameters are
listed in Table~\ref{table:margestats}, together with their 95\%
confidence ranges. They are compared to the parameters preferred by
\citet{Guo2012} for a fit of the previous model to the $z=0$ data
alone, assuming a WMAP7 cosmology. The most substantial differences
occur for the star formation efficiency, which is considerably higher
in the new model, and for the supernova feedback parameters, with high
reheating and ejection efficiencies in the new model extending to
higher mass systems. These changes combine with the new
reincorporation scheme to ensure that the number density of galaxies
below $M_{\star}$ is strongly suppressed at early times, but grows
through the return of ejected gas at later times so that the observed
abundance of these galaxies at $z=0$ is still reproduced. This later
formation increases the low-redshift star formation rates of low-mass
galaxies, making their colours bluer. In
Figure~\ref{fig:feedback_scalings} we present a graphical comparison
of the changes in the relevant efficiencies between G11 and this work.

In our new model the mass-loading of the winds which eject gas from
galaxy disks into their hot haloes depends less strongly on the
rotation velocity of the disks and is larger for all but the smallest
disks than in the G11 model. (This is $\epsilon_{\rm disk}$ in
equation~\ref{eq:reheat}; its scaling with $V_{\rm max}$ is
illustrated for the old and new models in the left panel of
Figure~\ref{fig:feedback_scalings}.) The values now required are similar
to those inferred from observational data both at low and at high
redshift \citep{Martin1999, Strickland2009, Steidel2010}.

As can be seen from the middle panel of
Figure~\ref{fig:feedback_scalings}, the values now required for the
parameters in equation~(\ref{eq:ejection2}) imply that all the
available energy from supernovae is used to reheat and eject gas in
typical disk galaxies (i.e. $\epsilon_{halo}=1$ for
$V_{\rm{max}}<300$~km/s). Arguably, this efficiency is
unsatisfactorily high since observations of supernova remnants suggest
that a significant fraction of their kinetic energy is thermalised and
radiated away. Nevertheless, we note that within our $2\sigma$ allowed
region $\epsilon_{halo}$ can be as low as 60\% for galaxies like the
Milky Way, and in any case the total energy available to drive winds
could exceed that we have assumed by a factor of two or
more. For example, we have considered only supernova feedback,
neglecting input due to radiation and winds from massive stars.

Finally, as already mentioned briefly above, the third panel of
Figure~\ref{fig:feedback_scalings} shows that the reincorporation
times for gas ejected from a halo depend more strongly on halo virial
velocity $V_{\rm vir}$ than in G11, and that at fixed $V_{\rm vir}$ the
dependence on redshift is reversed. Thus at $z=0$ reincorporation is
faster in the new model than in G11 for all galaxies with $V_{\rm
  vir}>30$~km/s, whereas at $z=3$ this is only true for $V_{\rm
  vir}>150$~km/s -- reincorporation is substantially slower at virial
velocities typical of dwarf galaxies.

\begin{figure*}
\centering
\includegraphics[scale=0.5]{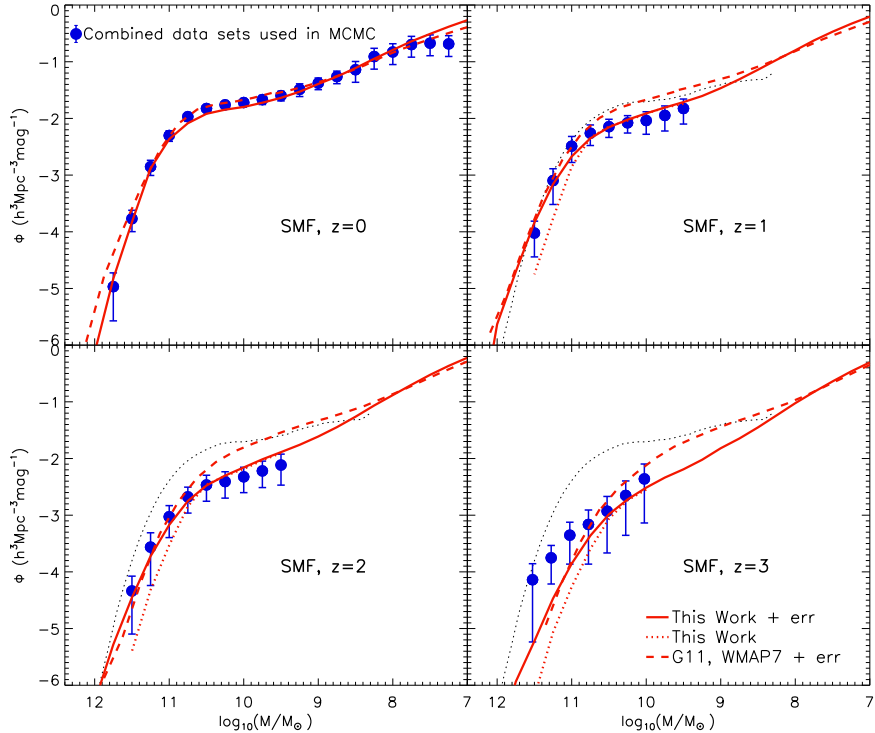}
\caption{Evolution of the stellar mass function from $z=3$ to
  $z=0$. Theoretical predictions from our new model and from
  \citet{Guo2012} are shown as solid and dashed red lines,
  respectively. The dotted red lines in the $z\geq 1$ panels are for
  the new model before convolution with a log-normal distribution of
  width 0.25 dex representing random uncertainties in the observed
  stellar masses. Blue dots with error bars show the observational
  constraints and their $1-\sigma$ uncertainties as adopted for the
  MCMC analysis. These result from combining a variety of datasets at
  each redshift. The individual data points are shown in
  Fig.~\ref{fig:smf_obs}. The $z=0$ results of \citet{Li2009} are
  repeated at all redshifts as a black dotted line. }
\label{fig:smf}
\end{figure*}

The increase in star formation efficiency (SFE) relative to the model
of \citet{Guo2012} primarily affects intermediate mass galaxies since
objects of lower masses have very short cooling times and star
formation rates are driven primarily by the infall of new gas and the
reincorporation of ejecta.  At high redshift, the
increased SFE compensates for the more efficient supernova feedback to
maintain the abundance of $M_{\star}$ objects. At lower redshift, the
effect is neutralized by the higher AGN radio mode efficiency in the
new model, so that the high mass cut-off in the mass function is
unchanged. The star formation efficiency parameter $\alpha_{\rm
  SF}$ regulates the conversion of cold gas into stars, so it is
important to note that, despite the new efficiency being $\sim 3$
times higher than in \citet{Guo2011} and $\sim 5$ times higher than in
\citet{Guo2012}, the predicted cold gas fractions of low-redshift
spiral and irregular galaxies are quite comparable. The higher
conversion rate from gas into stars is balanced by the increased
amount of gas available due to the delayed reincorporation of ejected
gas.  Finally, the new best-fit requires a higher value for the metal
yield than in G11.  This parameter is primarily constrained by the
relative abundances of bright galaxies in the $K$- and $B$-bands,
i.e. by the colours of massive elliptical galaxies.

\section{Consistency with the input constraints}

\label{sec:results}

In this section we compare the theoretical predictions of our new
model with the observational properties used as constraints in our
MCMC analysis: the stellar mass function, and the $K$- and
$B$-band luminosity functions from $z=3$ to $z=0$. All magnitudes are
AB, the stellar populations are \citet{Maraston2005} and the WMAP7
cosmology is adopted. For all properties we also compare with the
predictions of the G11 model tuned to the $z=0$ data alone, assuming
this same WMAP7 cosmology \citep{Guo2012} and re-calculated
luminosities for \citet{Maraston2005} stellar populations.

\subsection{Stellar mass functions}
\label{sec:smf}

\citet{Guo2011} found good agreement between the stellar mass function
of their model and the $z=0$ observations that they used to tune
its parameters.  However, an increasing excess of low-mass objects was
found when model predictions were compared to data at higher
redshifts.  This excess was not a consequence of the high value of
$\sigma_8$ adopted for the original Millennium
simulations, since it remained equally strong when the simulations
were scaled to a WMAP7 cosmology by \citet{Guo2012}. 

The observed number densities of massive and infrared-bright galaxies
are reproduced equally well in both cosmologies, once allowance is
made for the substantial random errors expected in the stellar masses
of high-redshift objects \citep{Henriques2012}. Within the G11 model
it appears impossible to retain this agreement, while fixing the
abundance evolution at lower mass.  Here we test how well our new
model for the reincorporation of ejected gas addresses this issue.
Note that we are attempting to build an {\it a priori} physical
simulation which reproduces the observed abundances of galaxies over
five orders of magnitude in stellar mass and over the last five sixths
of the age of the Universe.


\begin{figure*}
\centering
\includegraphics[scale=0.5]{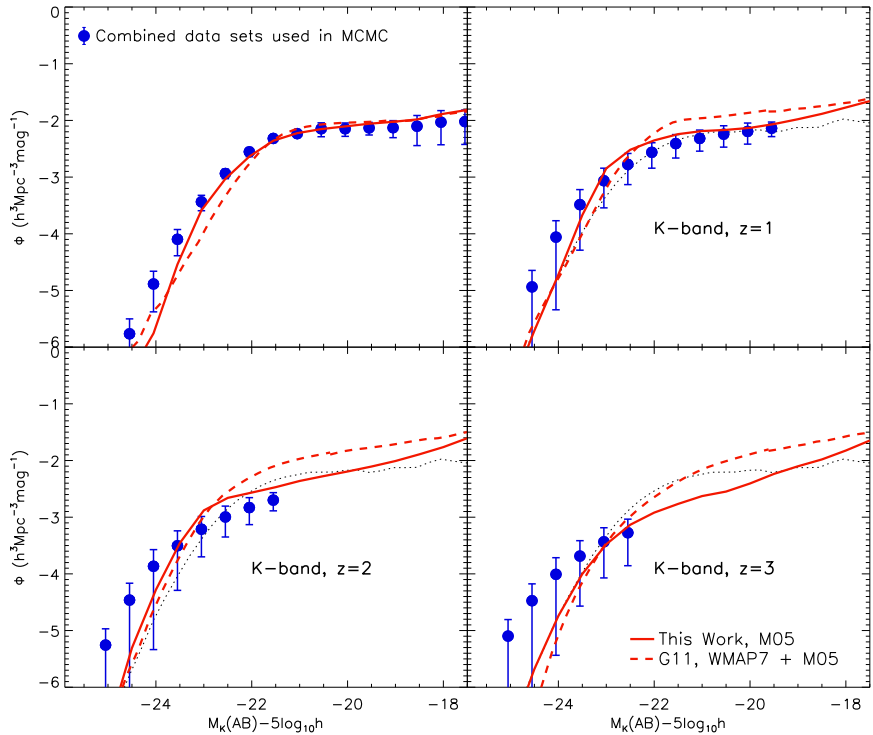}
\caption{Evolution of the rest-frame $K$-band luminosity function from
  $z=3$ to $z=0$. Results for our new model and for G11 are shown as
  solid and dashed red lines, respectively.  Blue dots with error bars
  show the observational constraints and their $1-\sigma$
  uncertainties as adopted for the MCMC analysis. These result from
  combining a variety of datasets at each redshift. The individual
  datasets are shown in Fig.~\ref{fig:kband_obs}. The $z=0$ results of
  \citet{Jones2006} are repeated at all redshifts as a black dotted
  line.}
\label{fig:kband}
\end{figure*}

In Fig.~\ref{fig:smf} we show the evolution of the stellar mass
function from $z=3$ (lower right panel) to $z=0$ (top left
panel). Results from G11 and from our new model with its best-fit
parameters are shown as dashed and solid red lines, respectively. The
dotted red line represents the new model before convolution with a
log-normal distribution of width 0.25 dex representing random
uncertainties in the observed stellar mass estimates. The shift
between dotted and solid red lines thus shows the predicted effect of
Eddington bias on the observed distribution. The models are compared
with the observational constraints adopted for our MCMC analysis,
which are derived from a comprehensive set of high-quality data. The
individual datasets are shown in Appendix \ref{app:obs_err}. 

The effect of modifying the treatment of gas reincorporation time is
clearly visible in Fig.~\ref{fig:smf}. The delayed return of gas
ejected in winds reduces the amount of gas available for star
formation at early times. The effect is pronounced in low-mass
galaxies, leading to a reduction in the number density at given
stellar mass by almost a factor of two at $z\geq2$.  The enhanced
reincorporation of ejected material at later times provides additional
fuel, allowing low-mass galaxies to form enough stars to continue to
match the observed abundances at $z=0$.

The model also matches the moderate but still noticeable evolution of
the abundance of massive galaxies. This agreement contrasts with
earlier work where the evolution in the observed abundances of high
mass galaxies appeared much weaker than predicted by hierarchical
models \citep[e.g.][]{Fontana2006, Caputi2006, Marchesini2009}. The
discrepancy can be explained by the combination of two effects.  For
high-redshift galaxies, photometric redshifts and stellar mass
estimates obtained by SED fitting have substantial uncertainties which
extend the high mass tail of the distribution of {\it estimated}
stellar masses. In addition, if masses are derived using population
synthesis models that omit the near-infrared emission expected from
intermediate age stars, the fluxes of high-redshift galaxies will be
interpreted as indicating overly high masses.  Once these two effects
are taken into account, both in correcting the data and in tuning the
model, quite reasonable agreement is obtained.


\begin{figure*}
\centering
\includegraphics[scale=0.5]{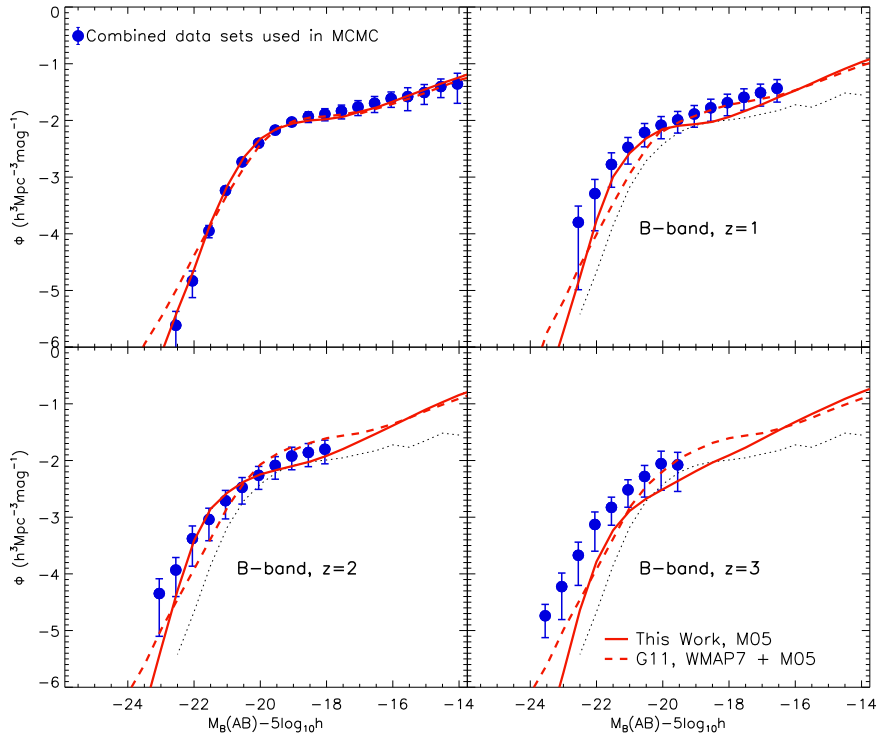}
\caption{Evolution of the rest-frame $B$-band luminosity function from
  $z=3$ to $z=0$. Blue dots with error bars show the observational
  constraints and their $1-\sigma$ uncertainties as adopted for the
  MCMC analysis. These result from combining a variety of datasets at
  each redshift.  The individual datasets are shown in
  Fig.~\ref{fig:bband_obs}. The $z=0$ results of \citet{Jones2006} are
  repeated at all redshifts as a black dotted line.}
\label{fig:bband}
\end{figure*}

\subsection{The rest-frame $K$-band luminosity function}
\label{sec:kband}

Most previous galaxy formation models have failed to match the
observed evolution of the faint end of the $K$-band luminosity
function \citep{Kitzbichler2007, Cirasuolo2010, Henriques2011,
  Henriques2012, Somerville2011}. This is a direct consequence of the
overly early build up of the dwarf galaxy population seen in the
stellar mass function. At the other extreme, model predictions for the
evolution in abundance of the brightest $K$-band galaxies have only
quite recently been reconciled with observations \citep{Bower2008,
  Tonini2010, Fontanot2010, Henriques2011, Henriques2012,
  Somerville2011}.

In Figure~\ref{fig:kband}, we compare the predicted evolution of the
$K$-band luminosity function out to redshift $z=3$ with the
observational data we used as constraints for our MCMC chains. Results
from our new model and from G11 are shown as solid and dashed red
lines, respectively, in both cases assuming \citet{Maraston2005}
stellar populations. The observational constraints are based on
combining results from a number of studies, which are shown
individually in Appendix \ref{app:obs_err}. Once the new
reincorporation prescription is included, our model agrees reasonably
well with the observational data over the full luminosity and redshift
ranges considered. The reduced number density of dwarfs at high
redshift caused by the delayed reincorporation of gas has a direct
impact on the abundances of faint near-infrared galaxies. This
corrects the previous excess of faint objects while maintaining
agreement at the bright end.

The most obvious remaining discrepancy is a residual overabundance of
faint galaxies at $z=2$. Within our present model this cannot be
corrected without worsening the good fit to the stellar mass and
$B$-band luminosity functions of the corresponding $z=2$ galaxies and
their $z=3$ progenitors. It is plausible that the problem stems from
residual systematic errors in one or more of these observational
datasets, but
it could also reflect a problem in the stellar population modelling of
$1\,{\rm Gyr}$ old populations.

Despite the considerable uncertainties that still remain in
theoretical predictions, it is interesting to note that the evolution
of the bright tail of the near-infrared luminosity function of
galaxies can be matched by a model that simultaneously reproduces the
observed evolution of the high-mass tail of the stellar mass function.
Until recently, stellar population synthesis models assumed
near-infrared emission to come predominantly from old populations. If
this were true, the weak evolution in the characteristic luminosity
$L_{\star}$ at rest-frame $K$ from $z=0$ to $z=3$, would imply
substantially less evolution of the characteristic mass of the
stellar mass function over this same redshift range than is expected
in hierarchical models of the kind we study here.  If, however,
intermediate age populations also emit significantly in the
near-infrared \citep{Maraston2005, Charlot2007, Conroy2009}, then the
observed luminosities of the high-redshift galaxies correspond to
significantly lower stellar masses, and the observed evolution of the
stellar mass and near-infrared luminosity functions can be matched
simultaneously. (This is already apparent from the combined results of
\citealt{Guo2011} and \citealt{Henriques2012}.) Acceptable results are
obtained for both the \citet{Maraston2005} and \citet{Charlot2007}
stellar population models, but using the older \citet{Bruzual2003}
models leads to a significant underprediction of $L_{\star}$ in the
$K$-band at higher redshift.


\subsection{The rest-frame $B$-band luminosity function}
\label{sec:bband}


Much of the light emitted in the $B$-band is produced by young stars
in star-forming regions. In addition, populations become bluer with
decreasing metallicity and so emit proportionately more in the
$B$-band.  Finally, star-forming galaxies also contain dust which
absorbs strongly at $B$. Thus, matching observations of the $B$-band
luminosity function and its evolution requires detailed modelling of
star formation, chemical enrichment and dust production. In our
semi-analytic model, star formation happens both quiescently and in
merger-induced bursts, and dust is modelled separately for the general
interstellar medium and for the birth clouds of young stars.  Dust
extinction is taken to be proportional to the column density of cold
gas in both cases \citep{DeLucia2007}.  Metals are returned to the
cold gas phase after each star formation episode and later follow the
gas as it is transferred between the various baryonic phases
\citep{Delucia2004b}.  \citet{Kitzbichler2007} found that although an
earlier version of the Munich model agreed well with the observed
$B$-band luminosity function at $z=0$, it significantly underpredicted
$B$-band luminosity functions at high redshift. They corrected this in
an {\it ad hoc} way by lowering the dust-to-cold gas ratios adopted at
early times. The evolution of this ratio with time was slightly modified by
\citet{Guo2009} and included in the G11 version of the Munich model,
leading to similar results for high-redshift luminosity functions
\citep{Henriques2012}, and it is also included here.

In Figure~\ref{fig:bband}, we compare predictions for $B$-band
luminosity functions from $z=3$ to $z=0$ with the observational
constraints used in our MCMC study. The individual datasets combined
to obtain these constraints are shown in Appendix
\ref{app:obs_err}. Results from our new model with its best-fit
parameters and from the original G11 model are shown as solid and
dashed red lines, respectively. At $z\sim 0$, the model is compared to
the $b_j$-band luminosity function of the 2dF survey assuming
$b_j=B-0.267(B-V)$ \citep{Norberg2002}, at higher redshift the
luminosity functions are for Johnson $B$. The delayed formation of
low-mass objects in our new model reduces the amplitude of the
high-redshift luminosity functions at fainter magnitudes, but the
large observational uncertainties mean that both models are consistent
with observation up to $z=2$. At $z=3$, both models underpredict the observed
abundance at all magnitudes, especially the brightest ones. Errors in
the treatments of star formation, enrichment and dust obscuration, as
well as uncertainties in the stellar population models, may all
contribute to this discrepancy.


\section{Model Predictions}

\label{sec:add_observations}
In this section we show model predictions for galaxy properties other
than those used as constraints in our MCMC sampling. These include
rest-frame colours, star formation rates, ages, stellar metallicities,
clustering, and the relation between halo mass and stellar mass. We
focus on low-redshift data, except that we analyse the evolution of
the halo-mass--stellar-mass relation from $z=3$ to $z=0$. In most
cases, we compare the predictions with data from the Sloan Digital Sky
Survey (SDSS) because the very large volume of the survey results in
small statistical and cosmic variance uncertainties. Our goal is to
investigate whether our improved representation of the evolution of
stellar mass and luminosity functions affects any of the other
discrepancies between the G11 model and observed galaxy
populations. We find that at $z=0$ low-mass galaxies in our new model
have higher star formation rates, younger stellar populations, are
more often blue and are less clustered on small scales than in the G11
model. In almost all cases this improves the agreement with the
observed SDSS galaxy population.  Since these properties were not used
as constraints when setting parameters, this should be viewed as a
significant success for the new model.

\subsection{Rest-frame colours}
\label{sec:colours}

\begin{figure*}
\centering
\includegraphics[width=17.9cm]{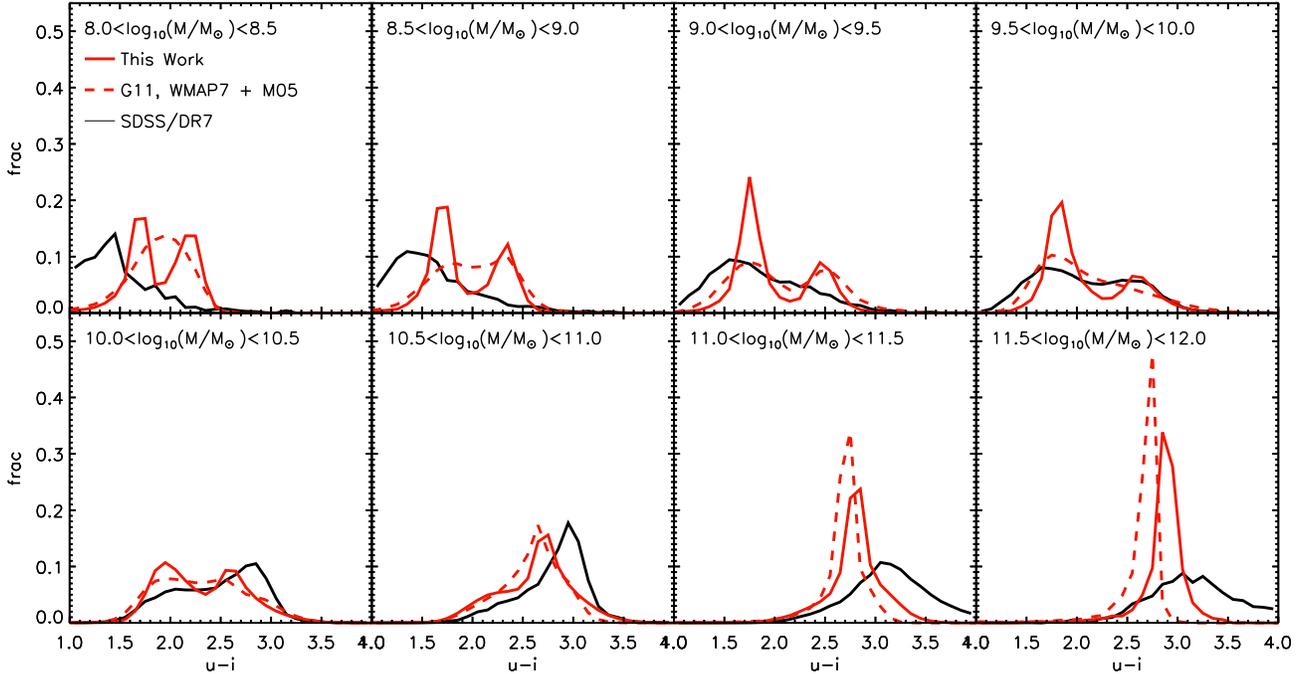}
\caption{Histograms of $u-i$ colour for $z=0$ galaxies in eight
  disjoint stellar mass ranges. Results for our new best-fit model
  (solid red lines) are compared with results from G11 (dashed red
  lines) and with low-redshift data from SDSS/DR7 (solid black
  lines). The models are based on the Millennium-II simulation for
  stellar masses below $10^{9.5}\Msun$ and on the Millennium
  simulation for higher masses.}
\label{fig:color_hist}
\end{figure*}

Galaxy colours at optical wavelengths are affected by a large number
of physical factors, making them difficult to model.  In particular,
current specific star formation rates, star formation and chemical
enrichment histories, the later stages of stellar evolution, and the
amount and distribution of dust can all have substantial effects on
optical colours. Despite this complexity, the observed population of
low-redshift galaxies separates fairly cleanly into a ``red sequence''
of relatively massive galaxies and a ``blue cloud'' of predominantly
low-mass dwarfs \citep{Kauffmann2003, Baldry2004, Weinmann2006a}.  In
the G11 model and its predecessors, the majority of dwarf galaxies are
much redder than observed, while massive galaxies are not red enough
\citep{Guo2011}.  Clearly, star formation is truncated too early in
the majority of model dwarfs, an effect which is little influenced by
switching from a WMAP1 to a WMAP7 cosmology \citep{Guo2012}. The
massive galaxies do form most of their stars at high redshift
\citep[e.g.][]{DeLucia2006} so their overly blue colours most likely
reflect a metallicity which is too low.

Figure~\ref{fig:color_hist} shows the distributions of $u-i$ colour
for eight stellar mass ranges from $10^8 \Msun$ to $10^{12}\Msun$ at
$z=0$. The new model with its best-fit parameters (solid red line) is
compared with the G11 model (dashed red line). These histograms use
data from the Millennium-II simulation for $10^8
\Msun<M_{\star}<10^{9.5}\Msun$ and from the Millennium simulation for
$10^{9.5}\Msun<M_{\star}<10^{12}\Msun$. As in \citet{Guo2011, Guo2012}
they are compared with data from SDSS/DR7, including $1/V_{\rm{max}}$
corrections so that the thin black histograms correspond to
volume-limited statistics. All histograms are normalized to have unit
integral.  There is a clear distinction in the observed distribution
between dwarfs, which are almost all blue, and massive galaxies, which
are almost all red. Only around the transition at
$M_*\sim10^{10.0}\Msun$ do both populations appear to exist in
comparable numbers. Our new best-fit model reproduces these trends
significantly better than the G11 model. The delayed reincorporation
of gas and hence the increased fuel supply at late times results in
the majority of low-mass galaxies continuing to form stars until
$z=0$. Despite this, the number of passive dwarf galaxies in the model
still substantially exceeds that observed.

This excess of passive dwarfs reflects, at least in part, the fact
that we impose a threshold for star formation (see
equations~\ref{eq:msamsfe}, \ref{eq:msamcriticaldensity} and
\ref{eq:msamcriticalmass}) since a substantial fraction of the red
low-mass galaxies have cold gas masses close to $m_{\rm crit}$. These
equations date back to \citet{Croton2006}, who based them on results
from \citet{Kennicutt1998} and \citet{Kauffmann1996a}, and they
clearly need updating to take account of recent improvements in our
observational understanding of the regularities underlying large-scale
star formation in galaxies \citep[e.g.][]{Bigiel2008}. First steps in
this direction have been made by \citet{Fu2012} who include a
prescription for atomic to molecular gas conversion, and assume star
formation to be proportional to the molecular content.  The observed
blue population of low-mass galaxies extends to significantly bluer
colours than predicted by the model. This appears to reflect the fact
that many of the observed systems have higher specific star-formation
rates than any star-forming dwarfs in the model (see
Fig.~\ref{fig:ssfr_age}).


For massive galaxies, the redder colours we obtain with our new model
are closer to those observed than are the colours from G11. As
discussed in the next section, this is a consequence of the higher
metal yield in our best-fit parameter set. This significantly
increases the metal abundance of these objects. In general, our
theoretical distributions are much narrower than the observed
distributions. This may reflect overly narrow metallicity
distributions in the model, dust effects or observational errors in
the observations.



\subsection{Stellar metallicities}
\label{sec:metals}

\begin{figure}
\centering
\includegraphics[width=8.6cm]{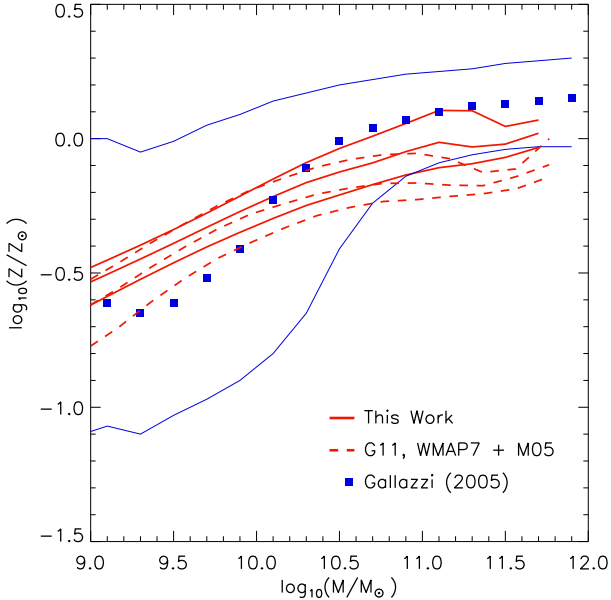}
\caption{The stellar mass-stellar metallicity relation at $z=0$. Our
  new model with its best-fit parameters (solid red lines) is compared
  with G11 (dashed red lines) and with SDSS data from
  \citet{Gallazzi2005} (blue squares and lines). In all cases, the
  central lines represent the median metallicity in each mass bin
  (blue squares for the observations), while the upper and lower lines
  represent the 16th and 84th percentiles of the distribution.}
\label{fig:metals}
\end{figure}

All chemical elements heavier than lithium are produced and deposited
in the gas phase during the later stages of stellar evolution. The
amount of metals in future generations of stars is then strongly
influenced by mixing processes and by exchanges of material between
the cold interstellar medium, the hot gas atmospheres of galaxies,
groups and clusters, and the circumgalactic medium. It follows that
changes in modelling of the physics of gas ejection and
reincorporation will be reflected in the stellar mass-metallicity
relation. In Fig.~\ref{fig:metals} we show this relation at $z=0$ for
our new model (solid red lines) for the model of G11 (dashed red
lines) and as estimated from SDSS data by \citet[][blue squares and
  lines]{Gallazzi2005}. For each sample, the median and the 16\%
and 84\% percentiles are shown as a function of stellar mass. Note
that \citet{Gallazzi2005} estimate the 1-$\sigma$ uncertainties in
their SDSS metallicity estimates to range from 0.6~dex at $10^9 \Msun$
to 0.2~dex above $10^{11}\Msun$. This can account for much but not all
of the difference in scatter between the models and the observations.
The optical to near-infrared colours of galaxies are strongly
influenced by their metal content, so the constraints we applied on
stellar mass and luminosity functions are reflected in the model
mass-metallicity relation. As a result, despite the differences in the
feedback physics, the predicted distribution shapes are very similar
in the two models. Both follow the observed trend approximately,
although with a significantly shallower slope.

A noticeable difference between the two models is the higher overall
metallicity in the new best-fit. This produces massive galaxies with
abundances closer to those observed, but also somewhat overpredicts the
metallicities of dwarfs. This is a consequence of the higher yield
preferred for the new model by our MCMC chains (0.047 as opposed to
0.03 in G11). The higher metallicity is responsible for the redder
colours of massive objects, as shown in the previous section, for the
increase in the number density of bright $K$-band galaxies, and for a
reduction in the abundance of bright $B$-band objects at $z=0$ (see
Sections~\ref{sec:kband} and \ref{sec:bband}, respectively).

If the uncertainty estimates of \citet{Gallazzi2005} are accurate, the
intrinsic scatter in the observed relation is considerably larger than
in the model. This might explain, at least in part, why the
red-sequence and blue-cloud populations show less scatter in colour in
the models than in the real data (Fig.~\ref{fig:color_hist}). The
steepness of the observed median relation is also puzzling,
particularly since the models seem to agree with the observed relation
between gas-phase metallicity and stellar mass (see G11 -- the current
model produces similar results). The model slopes in
Fig.~\ref{fig:metals} are quite similar to those found by other
published semi-analytic models \citep[e.g][]{DeLucia2012}. At the
massive end, a closer fit to observation might be obtained by more
effective tidal disruption of satellites \citep{Henriques2008,
  Henriques2010}; the high metallicities of stars formed {\it in situ}
are then less diluted by stars accreted through minor mergers because
these are instead assigned to the intracluster light.  We note,
however, that the G11 model already includes a substantial
intracluster light component in rich clusters. For stellar masses
below $\sim 10^{10}\Msun$, strong feedback from supernovae is crucial
in order to keep the metal abundances low \citep{Bertone2007,
  Henriques2010, DeLucia2012}.

\begin{figure}
\centering
\includegraphics[width=8.6cm]{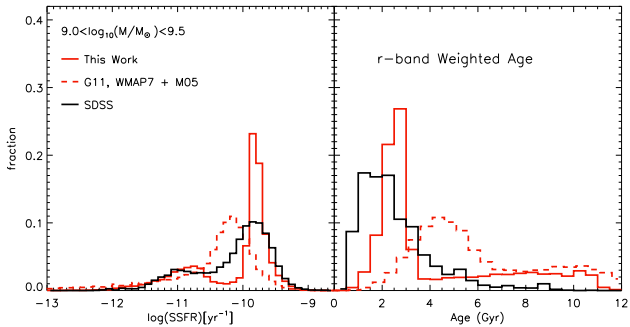}
\caption{The distribution of specific star formation rate (left panel)
  and $r$-band weighted age (right panel) for galaxies with
  $10^9\Msun<M_{\star}<10^{9.5}\Msun$. Predictions for our new model
  with its best-fit parameters are shown as solid red lines, while
  results from G11 are shown as dashed red lines (both are based on
  the Millennium-II simulation). The models are compared with
  SDSS data; specific star formation rate estimates come from
  \citet{Salim2007} and stellar ages from \citet{Gallazzi2005}. Both
  are shown as solid black lines.}
\label{fig:ssfr_age}
\end{figure}

\subsection{Ages and specific star formation rates in dwarf galaxies}
\label{sec:ages}

In a recent paper, \citet{Weinmann2012} analyzed the overly early
build-up of dwarf galaxies in semi-analytic models and in SPH
simulations by studying their ages and star formation rates at
$z=0$. As expected, the premature formation of dwarfs leads to older
ages and lower star formation rates than observed in the local
Universe, with similar discrepancies seen in the G11 model and in the
hydrodynamic simulations of \citet{Oppenheimer2008}.
\citet{Weinmann2012} focused on central galaxies, since including
satellites would require treatment of additional physical processes
and so introduce extra complications into the analysis. Nevertheless,
they do note that the disagreement between star formation properties
in the models and those observed is considerably worse for satellites
than for centrals, as already shown clearly by earlier work
\citep{Weinmann2009, Wang2012}. Using their own semi-analytic model,
\citet{Bower2012} found similar trends when looking at the number
density of low-mass galaxies at high redshift and their star formation
rates at $z=0$. Our new model correctly predicts the number density of
low-mass galaxies as a function of cosmic time, so it is interesting
to see if these discrepancies persist. In this section we compare
model predictions with data for all types of galaxies, including
satellites.

In Fig.~\ref{fig:ssfr_age}, we compare the distributions of specific
star formation rate (left panel) and $r$-band luminosity-weighted age
(right panel) for low-mass galaxies
($10^9\Msun<M_{\star}<10^{9.5}\Msun$) with observations at $z=0$. Our
new model with its best-fit parameters is represented by the solid red
lines while results from G11 are shown as dashed red lines. SDSS data
from the MPA-JHU
catalogue\footnote{http://www.mpa-garching.mpg.de/SDSS/} are plotted
as solid black lines after applying a $1/V_{\rm max}$ correction to
convert them to volume-limited statistics. The star formation rate
data are taken from SDSS-DR7 \citep{Salim2007} while the
luminosity-weighted ages are from SDSS-DR4
\citep{Gallazzi2005}. Stellar mass estimates following
\citet{Kauffmann2003} are used in both cases. The longer
reincorporation time-scales for ejected gas in our new model delay the
growth of low-mass galaxies and shift star-formation activity to
significantly lower redshifts. As can be seen in the left panel of
Fig.~\ref{fig:ssfr_age} this increases specific star formation rates
at $z=0$ by almost a factor of three and brings them into relatively
good agreement with those observed.

This later star formation results in substantially younger
luminosity-weighted ages. In the right panel of
Fig.~\ref{fig:ssfr_age}, the mode of the distribution for low-mass
galaxies drops by almost a factor of two to $\sim 2.5$ Gyr. This is in
considerably better agreement with SDSS data, although a substantial
further reduction still seems to be required. Note also the long tail
to ``old'' ages in the model histograms which is almost absent in the
observations. This corresponds to the red (primarily) satellite
population seen in the relevant panel of Fig.~\ref{fig:color_hist}. We
note that the age discrepancy for the blue population between the G11
model and the SDSS data appears larger here than in
\citet{Weinmann2012} because that paper compared $V$-band weighted
ages from the models with $r$-band weighted ages from SDSS.

\subsection{Stellar-mass -- halo-mass relation}
\label{sec:mstar_mvir}
The amount of cold gas, and hence of fuel for star formation, in a
galaxy depends directly on the mass of its halo. At early times the
ratio of baryons to dark matter is almost uniform on the scales which
build galaxies, and although later evolution modifies this
proportionality considerably, distributing the baryons differently
between the stars and the various gas components, it does not erase
the strong trend for more massive haloes to have more massive central
galaxies.  For this reason, a monotonic relation between the stellar
mass of a central galaxy and the mass of its halo is expected to be a
good first approximation to the outcome of the galaxy formation
process. Since the halo of a satellite galaxy can be strongly
truncated by tidal effects without a correspondingly large reduction
in the mass of its central galaxy, the stellar mass of a satellite is
expected to be more tightly related to the maximum past mass of its
(sub)halo than to its current subhalo mass (see  e.g. \citealt{Reddick2012}).

\begin{figure}
\centering
\includegraphics[width=8.6cm]{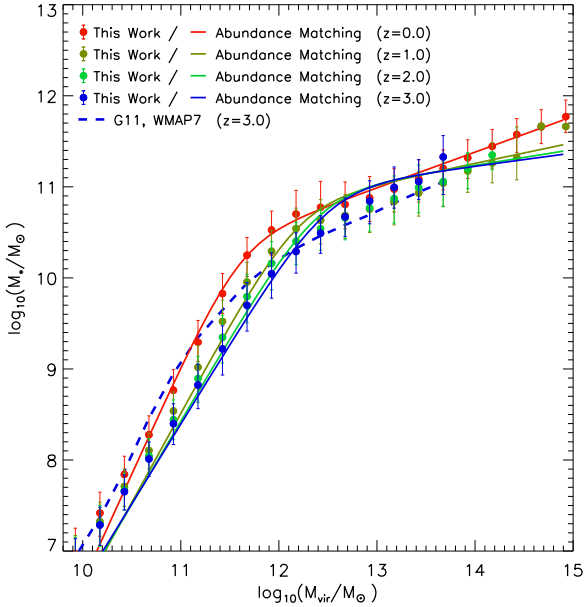}
\caption{Stellar mass as a function of maximum past halo mass for the
  galaxy populations present at redshifts 3, 2, 1 and 0. Filled
  circles show median values at each halo mass for our new model with
  its best-fit parameters, with error bars indicating the $\pm1\sigma$
  scatter.  For comparison, solid lines show the relations
  obtained at these same redshifts (coded by color) using abundance
  matching techniques by \citet{Moster2012}. The dashed blue line is
  the median relation at $z=3$ for the G11 model.}
\label{fig:mstar_mvir}
\end{figure}

Under the assumption of a monotonic relation between central galaxy
stellar mass and maximum past halo mass, this SM--HM relation can be
obtained explicitly by comparing the observed abundance of galaxies as
a function of stellar mass with the simulated abundance of subhaloes
as a function of past maximum mass \citep[see][for various forms of
  this abundance matching argument]{Frenk1988, Vale2004, Conroy2006,
  Behroozi2010, Moster2010, Guo2010, Reddick2012}. This assumes, of
course, that the real Universe conforms to the $\Lambda$CDM model,
and, in addition, that all real galaxies correspond to a subhalo which
survives until the relevant redshift in the dark matter simulation
being used. This latter assumption is seriously violated in the galaxy
formation models of G11 and this paper \citep[see, for example, Fig.~14
  in][]{Guo2011} although the violation does not dramatically affect
the SM--HM relation.

\begin{figure*}
\centering
\includegraphics[width=17.9cm]{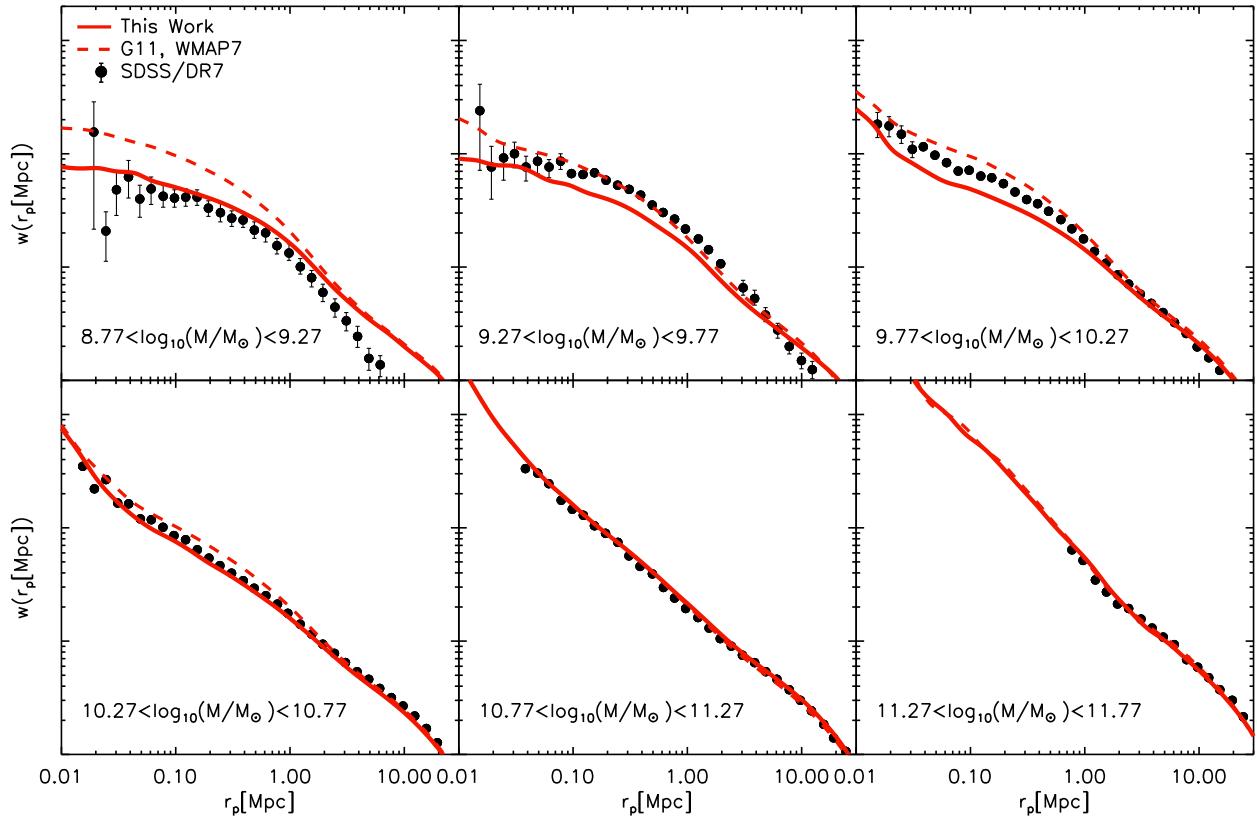}
\caption{Projected 2-point autocorrelation functions of galaxies in
  six disjoint stellar mass ranges. Results from our new model with
  its best-fit parameters (solid red lines) are compared with results
  from G11 (dashed red lines) and with observational data from
  SDSS/DR7 (solid black lines). The Millennium-II Simulation is used
  for stellar masses below $10^{9.77}\Msun$ and the Millennium
  Simulation for higher masses.}
\label{fig:correlation}
\end{figure*}

Although the (sub)halo abundance matching method appears very
successful at producing mock catalogues which match observation, it
has serious drawbacks. In the first place, since the observed
abundance of galaxies is used as {\it input} to the scheme, it cannot
be used to constrain the physics of galaxy formation. This has to be
done indirectly using the derived SM--HM relation, the clustering of
galaxies and their evolution. Secondly, the stellar masses of galaxies
undoubtedly do depend on aspects of their assembly history other than
the maximum past mass of their haloes, for example, whether they are
currently central or satellite galaxies, whether they recently
experienced a major merger, and so on.  These additional dependences
may bias mock catalogues constructed by abundance matching, and thus
the cosmological parameter values inferred by comparing their
large-scale structure to that observed. Such biases can only be
identified through models which explicitly follow galaxy formation.
Here we compare the SM--HM relation predicted by our new model to that
inferred by \citet{Moster2012} who applied an abundance matching
technique self-consistently across multiple redshifts using the
Millennium and Millennium-II simulations. Note that they corrected for
``missing'' subhaloes using the G11 model, rather than matching to the
subhaloes detected in the simulations alone.

Fig.~\ref{fig:mstar_mvir} compares results from \citet{Moster2012} for
redshifts 0, 1, 2 and 3 (solid lines) to the SM--HM
relation in our new model (filled circles give the median central
stellar mass at each halo mass, while ``error bars'' indicate the $\pm
1\sigma$ scatter).  Different colours encode redshift as indicated by
the label.  Model predictions are based on the Millennium Simulation
for $M_{\rm vir}>10^{11}\Msun$ and on the Millennium-II Simulation for
smaller masses. For comparison, we show the relation for the G11 model
at $z=3$ as a dotted blue line (at $z=0$ the G11 prediction is very
close to that of our new model).  For the new model, the agreement
with the SM--HM relation inferred directly by \citet{Moster2012} is
very good. This is, of course, expected since the new model is a good
fit to the high-redshift stellar mass functions we used as
constraints, and these are very similar to the observational results
used as input by \citet{Moster2012}. Notice that the characteristic
mass at which the SM--HM relation changes slope now shifts to higher
mass with increasing redshift, whereas it remained approximately
constant in the G11 model. As a result, stellar masses are
substantially lower in the new model at $z=3$ for haloes of mass
$M_{\rm vir}<10^{12}\Msun$ than in the G11 model, as required to
reduce the abundance of intermediate- and low-mass galaxies to the
observed level.  It is interesting that the $z=3$ relation lies below
the $z=0$ relation almost everywhere, showing that galaxy formation
was actually less efficient at given halo mass at $z=3$ than today,
despite the 64 times higher density within haloes at that time.

\subsection{Galaxy clustering}
\label{sec:clustering}
The spatial distribution of galaxies provides many important tests for
galaxy formation models. Clustering measurements hold information
about the cosmological parameters, the statistics of the initial
conditions, the nature of dark matter, the way in which galaxies grow
within dark matter haloes, the way in which they are modified by
environmental influences, and, on small scales, the rates of galaxy
collisions and mergers. The latter are of particular relevance for
semi-analytic models, since the limited resolution and the absence of
baryons in dark matter simulations cause subhaloes to be destroyed by
tides well before their central galaxies should themselves be
disrupted or merge into the central galaxy of their parent halo.  

The G11 model ties the position of each such orphan galaxy to that of
the dark matter particle which was most bound within its subhalo at
the last time this could be identified. At later times, the orphan's
position is a weighted average of those of the particle and of the
central galaxy of the main halo, coinciding with the particle at
subhalo disruption and with the central galaxy a dynamical friction
time later, at which point orphan and central galaxy merge.  This
scheme produces better agreement with observed small-scale galaxy
clustering than previous cruder models. G11 found it to reproduce the
autocorrelations of galaxies with $\log M_\star/\Msun > 10.8$ on all
scales larger than 30~kpc. However for smaller stellar masses they
found excess clustering on scales below 1~Mpc.  This discrepancy was
significantly smaller for a WMAP7 cosmology than for the original
WMAP1 cosmology, but remained noticeable.

In Figure~\ref{fig:correlation} we show projected autocorrelation
functions for galaxies in several disjoint stellar mass
ranges. Results for our new model with its best-fit parameters are
shown as solid red lines while results from G11 are shown as dashed
red lines. For both models, results from the Millennium-II Simulation
are plotted for $\log M_\star/\Msun < 9.77$ and from the Millennium
Simulation for higher masses. Model results are compared with the same
SDSS/DR7 data used in \citet{Guo2011} and \citet{Guo2012}. The new
model now accurately reproduces the observations for masses above
$10^{10.27}\Msun$ and for projected separations larger than 20~kpc. The
later assembly of stellar mass caused by our modified reincorporation
model reduces the masses of satellite galaxies relative to those of
centrals, because the satellites lose their ejecta reservoirs before
the gas can be reaccreted.  This weakens the ``1-halo'' contributions
to the autocorrelation functions and eliminates the small-scale
clustering excess for $8.77 < \log M_\star/\Msun < 9.27$ and
$10.27<\log M_\star/\Msun < 10.77$. In the two intermediate mass bins,
the reduction is actually too strong and the new model falls
significantly below the observations.  These results show that
small-scale clustering is quite sensitive to galaxy formation physics,
and as a result it is desirable to include clustering data among the
constraints when using MCMC sampling to explore galaxy formation
issues.



\section{Conclusions}

\label{sec:conclusions}

Semi-analytic galaxy formation models are designed to explore how the
many astrophysical processes which shape the formation and evolution
of galaxies are reflected in the systematic properties of the observed
galaxy population. In this paper we have extended the Monte Carlo
Markov Chain (MCMC) methodology introduced in this context by
\citet{Henriques2009} and \citet{Henriques2010} to allow population
properties at a wide range of redshifts to be used simultaneously as
constraints. This exploits one of the strong points of the
semi-analytic programme, the fact that its predictions at all
redshifts are, by construction, consistent both with the specific
physical assumptions of the model and with the growth of structure in
a $\Lambda$CDM universe. 

As observational input, we decided to use stellar mass functions and
rest-frame $B$- and $K$-band luminosity functions at $z=0$, 1, 2 and
3. We combined recent high quality observational determinations to
obtain a uniform set of constraints with plausible and consistent
estimates of their uncertainties. We adopted the modelling framework
of \citet{Guo2011} and used our optimised MCMC scheme to explore
allowed values for 11 of its parameters, those governing the physics
of star formation, black hole growth, feedback-driven gas flows and
chemical enrichment. 

When constrained using data at $z=0$ alone, the MCMC modelling gave
best-fit parameters close to those preferred by G11. Similarly good
fits were obtained when the data at any other single redshift were
used as constraints. However, no single parameter set could provide a
good fit at all four redshifts.  The allowed regions at different
epochs overlap for most parameters, but the timescale for
reincorporation of ejected gas is required to vary differently with
redshift than in the G11 model. This explains why this model, like
most other physically based galaxy population models \citep[see, for
example,][]{Fontanot2009, Cirasuolo2010, Somerville2011,
  Henriques2012, Guo2012, Bower2012, Weinmann2012} overpredicts the
abundance of lower mass galaxies at high redshift once parameters are
set to fit the present-day galaxy population.

In order to identify the modifications required to match galaxy
properties at all epochs, we introduced arbitrary power-law scalings
of ejecta return time with redshift and with halo virial velocity, and
we reran our MCMC sampling constrained by the abundance data at all
four redshifts together. This demonstrated that the extended model
could indeed fit the observed stellar mass and luminosity functions at
all four redshifts simultaneously. The ejecta return time was required
to scale inversely with halo virial mass and to be independent of
redshift. This simple result is quite similar to the scaling which
\citet{Oppenheimer2008} and \citet{Oppenheimer2010} found in their
cosmological hydrodynamic simulations, where the chosen feedback
recipes gave reasonable fits to the observed properties of
circumgalactic gas.

A comparison of our new best-fit model with the preferred model of G11
showed quite substantial changes to the efficiency and scaling of a
number of the star formation and feedback processes, despite the fact
that the two models produce very similar galaxy populations at
$z=0$. In the new model, the efficiency with which the gas disk makes
stars has increased substantially (by more than a factor of five),
supernova energy is used at maximum efficiency to eject gas in all but
the most massive galaxies, AGN feedback is more efficient (by more
than a factor of three), the mass-loading factor for disk winds
depends much more moderately on galaxy rotation velocity, and ejecta
return times scale more moderately with halo virial velocity $V_{\rm
  vir}$ but oppositely with redshift (at fixed $V_{\rm vir}$) than in
the G11 model. Combined the modifications ensure a suppression of the
formation of low-mass galaxies at early times and a more efficient
build-up at $z<2$, while maintaining the previous evolution of the
number density of the most massive objects. An increased yield leads
to better reproduction of the metallicities of massive galaxies.

The fact that models with such substantial differences produce similar
present-day populations is a dramatic illustration of the
interconnectedness of the many processes influencing galaxy formation.
Nevertheless, our MCMC analysis demonstrates that there are no
significant degeneracies when each model is fit to the observational
data it was designed to interpret.  All the parameters are required
for an adequate fit. The physical interest of the fits lies in the
specific scalings and efficiencies implied, in particular, in checking
whether the efficiencies lie within their plausible ranges. In this
context it may be worrying that our new model requires supernova
energy to be used so efficiently to eject gas from galaxy disks.

In our new model ejected gas returns to lower mass haloes much less
rapidly at high redshift than in the G11 model. As a result, low mass
galaxies grow more slowly, and their abundance as a function of
stellar mass and luminosity is a better fit to the observations. At
$z<1$, however, ejecta return to all but the smallest haloes more
rapidly than in the earlier model, leading to stronger late-time star
formation and similar $z=0$ stellar masses. By construction, these
changes eliminate one of the main problems which \citet{Guo2012}
identified in their model: its overly early production of lower mass
galaxies. It is interesting that the other two problems mentioned by
the authors are also reduced, even though the relevant observations
were not used as constraints in our MCMC modelling. As a result, in
the new model:
\begin{itemize}
\item The observed galaxy abundances are correctly represented over 5
  order of magnitude in mass ($7<\log M_\star/\Msun <12$) and
  throughout most of the age of the Universe.
\item The fraction of dwarf galaxies ($\log M_\star/\Msun <9.5$) which
  are blue, young and actively star-forming at $z=0$ is substantially
  higher than in the model of \citet{Guo2012}.
\item The clustering of dwarf galaxies  is reduced as result of the
  increased central to satellite ratio at fixed stellar mass, and is
  no longer systematically above that observed.
\end{itemize}
Nevertheless, at the lowest masses, star-forming galaxies are still
older and the passive fraction is still significantly larger than
observed. This problem may be related to the star-formation threshold
adopted in the model (see Appendix A1), since many of the dwarfs,
while containing considerable gas, are sitting very close to the
threshold.

The increased heavy element yield in the new model produces higher
metallicities and redder colours for massive galaxies, improving the
fit to observations.  On the other hand, the slope of the stellar
mass-metallicity relation remains shallower than observed, so the
stellar metallicities of low-mass galaxies are now
overpredicted. \citet{Henriques2010} noted that the slope of this
relation can be increased by allowing more efficient tidal disruption
of satellites. The stellar populations of massive galaxies are then
less diluted by accretion of low-metallicity material from merging
satellites.  It remains to be seen whether this can improve the
chemical properties of the model without destroying its ability to fit
the luminosities of massive galaxies which currently grow
substantially more through merging than through star formation
\citep[see][]{Guo2008}.  A more complete treatment of chemical
evolution, allowing detailed comparison with the observational
information now available on stellar and gas-phase abundances and
abundance ratios, will undoubtedly be a fruitful extension of the kind
of galaxy population modelling presented in this paper.

\section*{Acknowledgements}

The computations used in this paper were performed on the Cosmology
Machine supercomputer at the Institute for Computational Cosmology,
which is part of the DiRAC Facility jointly funded by STFC, the Large
Facilities Capital Fund of BIS, and Durham University.  The work of
BH, SW, RA and GL was supported by Advanced Grant 246797 ``GALFORMOD''
from the European Research Council. PT was supported by the Science
and Technology Facilities Council [grant number ST/I000976/1]. GQ
acknowledges support from the National basic research program of China
(973 program under grant No. 2009CB24901), the Young Researcher Grant
of National Astronomical Observatories, CAS, the NSFC grants program
(No. 11143005), as well as the Partner Group program of the Max Planck
Society. VS acknowledges support by the Deutsche
Forschungsgemeinschaft through Transregio 33, ``The Dark Universe''.
The authors thank Michele Cirasuolo, Danilo Marchesini and Rob Yates
for providing their observational data.

\bibliographystyle{mn2e} \bibliography{paper}

\appendix
\section{Physical Prescriptions}
\label{app:physics}

Our semi-analytic model is built on top of high-resolution N-body
simulations that provide a detailed model for the evolution of the
dark matter distribution.  At each time, a friends-of-friends
algorithm is used to define a set of haloes, each of which is then
further divided into a set of subhaloes, disjoint, self-bound particle
clumps with a single gravitational potential minimum where a galaxy is
assumed to form. Each subhalo at time $i$ is linked to a unique
descendant at time $i+1$ by following its constituent particles.
These links produce merger trees which specify the full assembly
history of every $z=0$ subhalo, and are the data structures on which
the semi-analytic model operates to follow the formation and evolution
of the galaxy that each subhalo contains. As subhaloes form and grow
through accretion, new dark matter is assumed to bring the global mean
baryon fraction with it. These baryons are transferred among the
various baryonic components associated with the subhalo (hot gas
atmosphere, cold disk gas, ejected gas, disk and bulge stars) by the
semi-analytic model.  In very low-mass haloes, gas infall is suppressed
by UV heating once the Universe is ionized. At higher mass, accretion
of newly acquired gas occurs in two modes: at high redshift and for
subhaloes with mass below $\sim 10^{11.5}\Msun$, cooling is rapid and
infalling gas accretes directly onto the central galaxy; at later
times and in more massive haloes, the gas shocks to form a quasi-static
atmosphere from which it gradually accretes onto the galaxy through a
cooling flow. Although this spherical model does not allow filamentary
inflow to coexist with a hot atmosphere, it produces galaxy accretion
rates as a function of halo mass and redshift which agree reasonably
with those derived from detailed gas dynamical simulations. In
practice, the situation is considerably complicated by the strong
galactic winds required in all realistic models.

Star formation occurs in cold disk gas, either quiescently or in
merger-induced bursts. A fraction of the stars formed (those with
masses above $8 \Msun$) is assumed to explode as type II
supernovae. This mass is instantaneously returned to the cold phase
and the energy released is used to reheat gas from the cold to the hot
phase. Left-over energy is used to eject gas from the hot phase into
an external reservoir. This ejected gas can later be reincorporated in
the hot phase, and it is this aspect of the G11 model which we modify
in this paper, as described in Section~\ref{sec:new_model}. The
evolution of massive stars also produces new metals that are deposited
in the cold phase. Each newly formed galaxy is assigned a central seed
black hole that then grows through accretion of cold gas during
mergers and through quiescent accretion of hot gas. When the black
hole reaches a sufficiently large mass, this quiescent accretion is
assumed to power a radio source which offsets cooling in the hot gas,
thereby quenching star formation.

In the following subsections, we present the equations describing
those aspects of the semi-analytic modelling for which the parameters
have been allowed to vary in our MCMC analysis. These parameters are
$\alpha_{\rm{SF}}$, $k_{\rm{AGN}}$, $f_{\rm{BH}}$, $\epsilon$,
$V_{\rm{reheat}}$, $\beta_{\rm{1}}$, $\eta$, $V_{\rm{eject}}$,
$\beta_{\rm{2}}$, $\gamma'$ and $Z_{\rm{yield}}$. For a more complete
description of the model we refer the reader to \citet{Guo2011}.


\subsection{Star formation}

Over most of the lifetime of a galaxy, star formation happens in a
quiescent mode through the fragmentation of cold gas that reaches a
sufficiently high density. The star formation rate in this mode is
modeled as
\begin{equation} \label{eq:msamsfe}
\dot{m}_{\star}
= \alpha_{\rm SF}\frac{(m_{\rm cold}-m_{\rm crit})}{t_{\rm dyn,disk}},
\end{equation}
where $m_{\rm cold}$ is the mass of cold gas and
$t_{\rm{dyn,disk}}=r_{\rm{disk}}/V_{\rm{max}}$ is the dynamical time
of the disk. Note that the mass locked up in stars over a time interval ${\rm d}t$ is
$(1-{\rm{R}})\dot{m}_{\star}\,{\rm d}t$, the rest being instantaneously returned to the 
cold disk gas. ($R=0.43$  here
denotes the recycled fraction as determined from the
\citet{Chabrier2003} initial mass function.) The threshold gas mass
above which star formation is assumed to occur, $m_{\rm{crit}}$, is derived from
a threshold surface density at each radius as given by \citet{Kauffmann1996a}:
\begin{equation} \label{eq:msamcriticaldensity}
\Sigma_{\rm{crit}}(R)=120\left(\frac{V_{\rm{vir}}}{200\,\rm{km\, s}^{-1}}\right)\left(\frac{R}{\rm{kpc}}\right)^{-1}\Msun \rm{pc}^{-2},
\end{equation}
leading to the estimate
\begin{equation} \label{eq:msamcriticalmass}
m_{\rm{crit}}=3.8\times 10^9\left(\frac{V_{\rm{vir}}}{200\,\rm{km\, s}^{-1}}\right)
\left(\frac{r_{\rm{disk}}}{10\,\rm{kpc}}\right)\Msun,
\end{equation}
where $r_{\mathrm{disk}}$ is obtained from the specific angular
momentum of the cold gas disk which in turn is calculated by summing
the contributions to the angular momentum vector from all infalling
gas \citep{Guo2011}.

During mergers it is assumed that the cold gas is disturbed
significantly, allowing a large fraction to reach the densities
necessary for fragmentation. This phenomenon is modeled \citep[following][]{Somerville2001}
as a higher
efficiency for star formation during merger events:
 \begin{equation} \label{eq:msamsfeburst}
\dot{m}_{\star}^{\rm{burst}}
= 0.56 \left(\frac{m_{\rm{sat}}}{m_{\rm{central}}}\right)^{0.7}m_{\rm{gas}}.
\end{equation}
The burst parameters are kept fixed in the MCMC sampling; of the parameters in this section,
only $\alpha_{\rm SF}$ is allowed to vary.

\subsection{Supernova feedback}
Massive stars release large amounts of energy into the surrounding
medium both through their stellar winds and when they explode as
supernovae.  This energy can reheat cold disk gas and inject it into
the hot atmosphere, and it may also eject hot gas from the virialised
regions of the galaxy/subhalo system. The ejecta may then fall back
into the system at some later time. This feedback-induced cycle is
treated by the semi-analytic model as a three stage process involving
gas reheating, gas ejection and the reincorporation of ejected gas.

The mass of gas reheated by star formation is assumed to be
directly proportional to the amount of stars formed:
\begin{equation} \label{eq:reheat}
\ \Delta m_{\rm reheated}=\epsilon_{\rm disk}\Delta m_{\star},
\end{equation}
where
\begin{equation} \label{eq:reheat2}
\ \epsilon_{\rm disk}=\epsilon \times \left[0.5+\left(\frac{V_{\rm max}}{V_{\rm reheat}}\right)^{-\beta_{1}}  \right].
\end{equation}
Any supernova energy left over after reheating the disk gas 
is used to eject material from the hot atmosphere into an external
reservoir. The total energy input from SN is assumed to be:
\begin{equation} \label{eq:ejection1}
\Delta E_{\rm SN}=\epsilon_{\rm halo}\times\frac{1}{2}\Delta
m_{\star}V_{\rm SN}^2,
\end{equation}
where
\begin{equation} \label{eq:ejection2}
\epsilon_{\rm halo} =\eta\times \left[0.5+\left(\frac{V_{\rm
        max}}{V_{\rm eject}}\right)^{-\beta_{2}}  \right] .
\end{equation}
The ejected mass is then derived from the residual energy after reheating:
\begin{equation} \label{eq:ejection3}
\frac{1}{2}\Delta M_{\rm eject}V_{\rm vir}^2=\Delta E_{\rm SN}-\frac{1}{2}\Delta M_{\rm reheat}V_{\rm vir}^2.
\end{equation}
As described in \citet{Guo2011}, if this equation gives $\delta
M_{\rm{ejec}}<0$, the reheated mass is assumed to saturate at $\delta
M_{\rm{reheat}}=\delta
E_{\rm{SN}}/\left(\frac{1}{2}V^2_{\rm{vir}}\right)$. Also,
$\epsilon_{\rm{halo}}$ is forced never to exceed unity. Both
conditions ensure that the total amount of energy used in feedback is
limited to that available from supernovae.  Material ejected from the hot
atmosphere can be reincorporated at later times according to the
models described in Section~\ref{sec:new_model}.

\subsection{AGN feedback}
Three processes are assumed to control the growth of massive central
black holes and their effects on their environment.  Cold gas
accretion during mergers fuels ``quasar mode'' growth which is the
principal route by which the black hole population gains mass. Black
holes are also assumed to accrete quiescently from hot gas atmospheres
in a ``radio mode'' which does not add significantly to their mass but
pumps energy into the hot gas, thus counteracting its cooling and
quenching the supply of new cold gas for star formation.  Finally
central black holes are assumed to merge when their parent galaxies
merge.

The amount of gas accreted in quasar mode is taken to be proportional
to the mass ratio between the two merging galaxies and to the their
total amount of cold gas:
\begin{equation} \label{eq:msamfbh}
\Delta m_{\rm BH,Q}=\frac{f_{\rm BH}(m_{\rm sat}/m_{\rm
    central})\,m_{\rm cold}}{1+(280\,\mathrm{km\,s}^{-1}/V_{\rm vir})^2}.
\end{equation}

The rate of radio mode accretion from the hot gas is taken to be
\begin{equation} \label{eq:msamkagn}
\dot{m}_{\rm BH,R}=k_{\rm AGN}\left(\frac{m_{\rm
    BH}}{10^8\,\Msun}\right)\left(\frac{f_{\rm
    hot}}{0.1}\right)\left(\frac{V_{\rm
    vir}}{200\,\mathrm{km\,s}^{-1}}\right)^3,
\end{equation}
where $m_{\rm {BH}}$ is the black hole mass and $f_{\rm{hot}}$ is the
mass fraction of hot gas in the halo. This accretion is assumed to
pump energy back into the hot gas through
mechanical heating at a substantial fraction of the Eddington rate:
\begin{equation} \label{eq:eddington}
L_{BH}=\zeta \dot{m}_{\rm{BH,R}}c^2,
\end{equation}
where $c$ is the speed of light and $\zeta$ is set to 0.1 (which in
combination with $k_{\rm{AGN}}$ determines the heating rate). This
energy is used to offset the cooling of gas from the hot atmosphere
onto the cold disk, resulting in a modified cooling rate,
\begin{equation} \label{eq:modcooling}
\dot{m}'_{\rm{cool}}=\dot{m}_{\rm{cool}}- \frac{L_{\rm{BH}}}{\frac{1}{2}V^2_{\rm{vir}}},
\end{equation}
which is required to be non-negative.

\subsection{Metal enrichment}

The metals produced by the stars are assumed to follow the same
processing cycle as the gas.  For every solar mass of stars formed, a
mass $Z_{\rm yield}\Msun$ of heavy elements is returned
instantaneously to the cold phase, where $Z_{\rm yield}$ is known as
the yield and is the only chemical enrichment parameter which we allow
to vary in our MCMC analysis. These heavy elements then follow the gas
flow. They can be locked in new long-lived stars, or they can be
reheated into the hot phase, ejected into the external reservoir,
reincorporated in the hot phase, and returned to the cold phase by
cooling. Each of the three gas phases is assumed to be chemically
homogeneous at each time, i.e. mixing of heavy elements within each
phase is assumed to be instantaneous. 

%
%

\section{Monte Carlo Markov Chain analysis}
\label{app:sampling}

As discussed in \citet{Henriques2009}, implementing the MCMC approach
for cosmological galaxy formation modelling on simulations like the MS
and MS-II raises significant challenges with respect to computational
power. In a brute force approach, the semi-analytic model would need
to follow the formation and evolution of up to 20 million galaxies
throughout the simulation volume for every parameter set proposed at a
MCMC step.  Population distributions of interest could then be
compared with the observational constraints to compute a likelihood.
The size of the calculation and the required number of steps make this
a very costly proposition. \citet{Henriques2009} and
\citet{Henriques2010} therefore restricted themselves to a small but
approximately representative sub-volume, just $1/512$ of the full
Millennium Simulation. While reasonably efficient, their approach did
not optimise the fidelity of the representation of the full model for
given computational resources. Here, we adopt a scheme which defines a
representative set of subhalo merger trees such that abundance
statistics such as those used as constraints in this paper can be
modelled with controlled and approximately uniform precision and at
minimum computational cost across the full range of stellar masses (or
luminosities) considered. It is also important to ensure that we have
a properly representative model at all the redshifts considered. Our 
scheme allows merger trees from the MS and MS-II to be used
simultaneously, significantly increasing the dynamical range in galaxy
mass/luminosity which can be used to constrain the model.

Consider the problem of estimating the galaxy luminosity function
$\Phi(L)$ of a specific model, which we take to be the function
obtained when the model is applied to the full MS and MS-II
simulations.  We want to find a small subset of the halo merger trees
which reproduces $\Phi(L)$ to some desired accuracy over some
luminosity range of interest.  Let us assume that our simulated volume
contains $N$ haloes, partitioned into a total of $I$ mass bins
labelled by $i$.  Then
\begin{equation}
N = \sum_{i=1}^I N_i,
\end{equation}
where $N_i$ is the number of haloes in the $i$'th mass bin. Further,
assume that we wish to estimate a luminosity function defined on a set
of $J$ luminosity bins labelled by the index $j$.  We can write the
luminosity function of the simulated volume as a whole as
\begin{equation}
\Phi_j = \sum_{i=1}^I N_i~\phi_{ij},
\end{equation}
where $\phi_{ij}$ is the average number of galaxies in luminosity
bin $j$ for a randomly selected simulation halo in mass bin $i$.
We seek a set of numbers $n_i$ such that choosing $n_i$ haloes at random
among the $N_i$ in the $i$th bin is sufficient to estimate the luminosity
to function to some preassigned accuracy, for example, $\pm 5\%$.

For a specific subsample of $n_i$ haloes in each bin $i$, a
straightforward estimate of $\Phi_j$ is
\begin{equation}
\label{eq:LFestimate}
\tilde{\Phi}_j = \sum_{i=1}^I\frac{N_i}{n_i}\sum_{k=1}^{n_i}s_{ijk},
\end{equation}
where $s_{ijk}$ is the number of galaxies in the $j$'th luminosity bin
for the $k$'th of the $n_i$ haloes in mass bin $i$. For different $k$
the $s_{ijk}$ are independent, identically distributed random
variables with non-negative integer values and mean $\phi_{ij}$. We
will assume that they are Poisson distributed, but we note that this,
in fact, overestimates their variance because each halo
contains one and only one central galaxy and such central galaxies dominate
the bright end of the luminosity function (see, for example, Fig.~24
of \citet{Guo2011}). With this Poisson assumption we have
\begin{equation}
\langle\tilde{\Phi}_j\rangle = \Phi_j
\end{equation}
and
\begin{equation}
{\rm var}\big[\tilde{\Phi}_j\big]=\sum_{i=1}^I\frac{N_i^2}{n_i}\phi_{ij}.
\end{equation}
If we wish the {\it rms} uncertainty of our luminosity function estimate
to be a fraction $F$ of its true value in each of the luminosity bins, we
then require
\begin{equation}
\label{eq:optimum}
\sum_{i=1}^I\frac{N_i^2}{n_i}\phi_{ij} = F^2~\Phi_j^2;~~~j=1,J.
\end{equation}

For a specific galaxy formation model implemented on the Millennium simulations,
the G11 model,  for example, we know the $N_i$, the $\Phi_j$ and the $\phi_{ij}$,
so equations~\ref{eq:optimum} are a set of linear constraint equations on
$F^2n_i$. In the specific case $I=J$, the matrix $\phi_{ij}$ is square and can
be inverted to get an explicit solution for the $n_i$,
\begin{equation}
\frac{1}{n_i} =
\frac{F^2}{N_i^2}\sum_{j=1}^J\Phi_j^2\phi^{-1}_{ji};~~~i=1,I.
\end{equation}
If $I>J$ and if for every $i$ at least one of the $\phi_{ij}$ is
non-zero, the solution space of equations~\ref{eq:optimum} is
$(I-J)$-dimensional and one can normally pick a set of $n_i$ values
satisfying the requirements $0\leq n_i\leq N_i$ and minimising some
computational cost function, for example, the total number of trees
chosen, $\sum_i n_i$, or the total number of galaxies modeled,
$\sum_{ij} n_i\phi_{ij}$. For our analysis we choose $F=0.05$ to
ensure that the uncertainty of the luminosity and stellar mass
functions estimated for our models are smaller than the observational
uncertainties on the data we compare them with. In total, fewer than
$1\%$ of all Millennium trees are needed to obtain estimates
consistent with the full set to this level of accuracy. In practice,
we use Millennium-II trees to calculate abundances for galaxies with
$\log M_\star/\Msun < 9.5$ and Millennium trees to calculate abundances
for more massive galaxies.

A representative and sufficiently large sample of haloes is needed at
all redshifts at which the model will be compared with observation.
Starting at $z=0$, we select the ``optimum'' number $n_i$ of FoF
haloes at random in each mass bin. When weighted by $N_i/n_i$, these
provide a representative sample of the full cosmic halo
population. Hence, when combined with the same weights, the galaxies
they contain at $z=0$ reproduce the present-day luminosity and stellar
mass functions (equation~\ref{eq:LFestimate}) and the galaxies
contained by their progenitors at $z=1$ (say) reproduce these same
functions at that time. It is unclear, however, whether there will be
enough progenitor haloes in each mass bin to satisfy our S/N
requirements at $z=1$. We check explicitly whether this is the case. If
not, we select sufficient additional $z=1$ haloes to restore the
required precision. Their formation trees are then followed back
together with those of the original haloes to $z=2$ and the process is
repeated. Note that the representative halo set and the associated
trees and weights are determined once using a fiducial model.  Galaxy
formation is then followed sequentially from high to low redshift in
this predefined set of trees for every parameter set in our MCMC
analysis. We have verified explicitly that these procedures produce
luminosity and stellar mass function estimates at all redshifts which
reproduce the functions for the full simulations at least as
accurately as inferred from the simplified analysis above.

\section{Observational errors}
\label{app:obs_err}

\begin{figure*}
\centering
\includegraphics[scale=0.5]{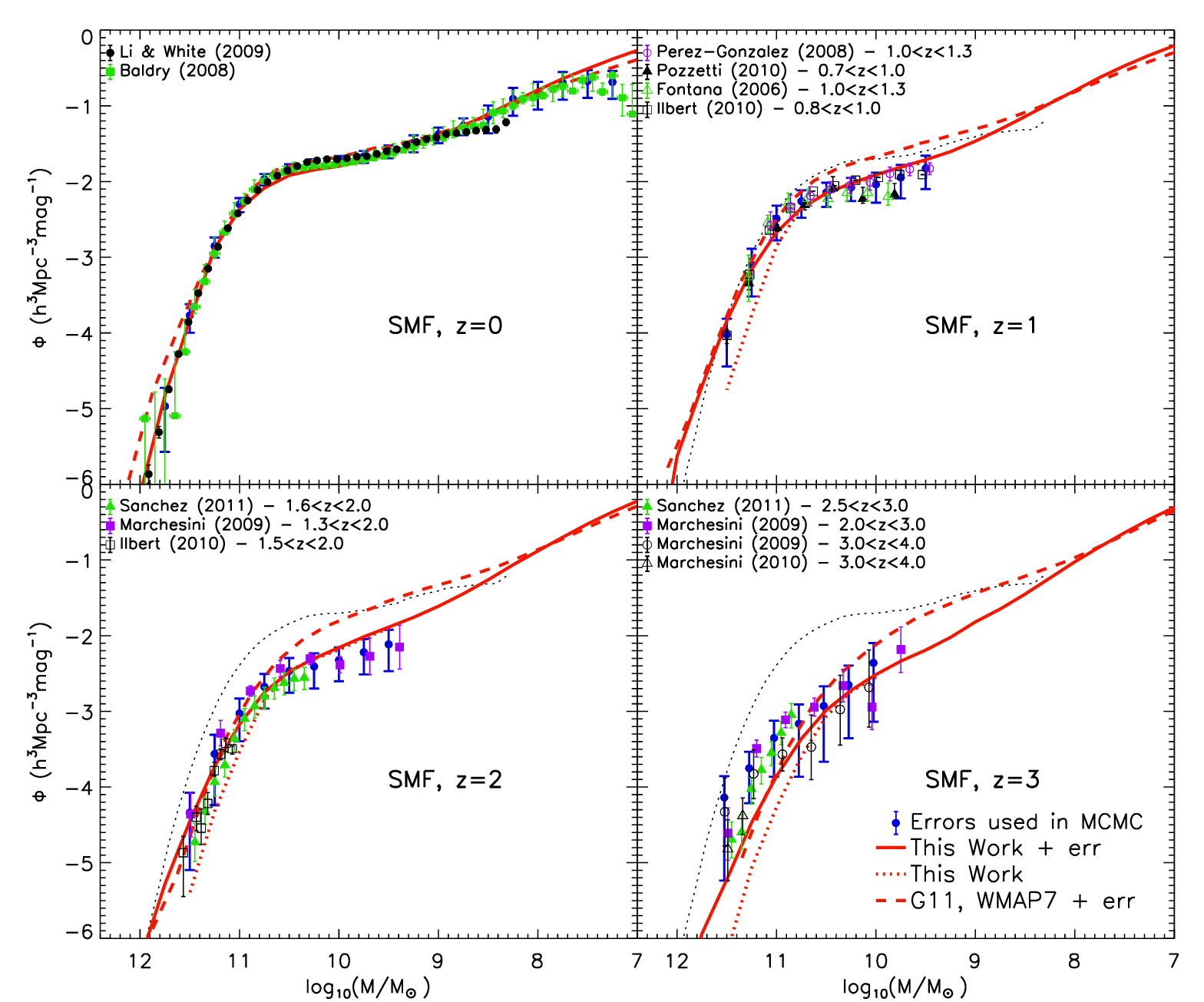}
\caption{Evolution of the stellar mass function from $z=3$ to $z=0$ as
  in Fig.~\ref{fig:smf}, except with the data points for the
  individual underlying surveys also shown. These surveys are
  \citet{Baldry2008} and \citet{Li2009} at $z=0$, and GOODS-MUSIC
  \citep{Fontana2006}, Spitzer \citep{Perez2008},
  \citet{Marchesini2009}, Spitzer-COSMOS \citep{Ilbert2010}, NEWFIRM
  \citep{Marchesini2010}, zCOSMOS \citep{Pozzetti2010} and COSMOS
  \citep{Sanchez2011} at higher redshift. All mass estimates at $z>0$,
  except \citet{Sanchez2011} have been shifted by -0.14 dex to convert
  from \citet{Bruzual2003} to \citet{Maraston2005} stellar populations
  \citep{Sanchez2011}. The $z=0$ results of \citet{Li2009} are repeated
  at all redshifts as a black dotted line.}
\label{fig:smf_obs}
\end{figure*}

\begin{figure*}
\centering
\includegraphics[scale=0.5]{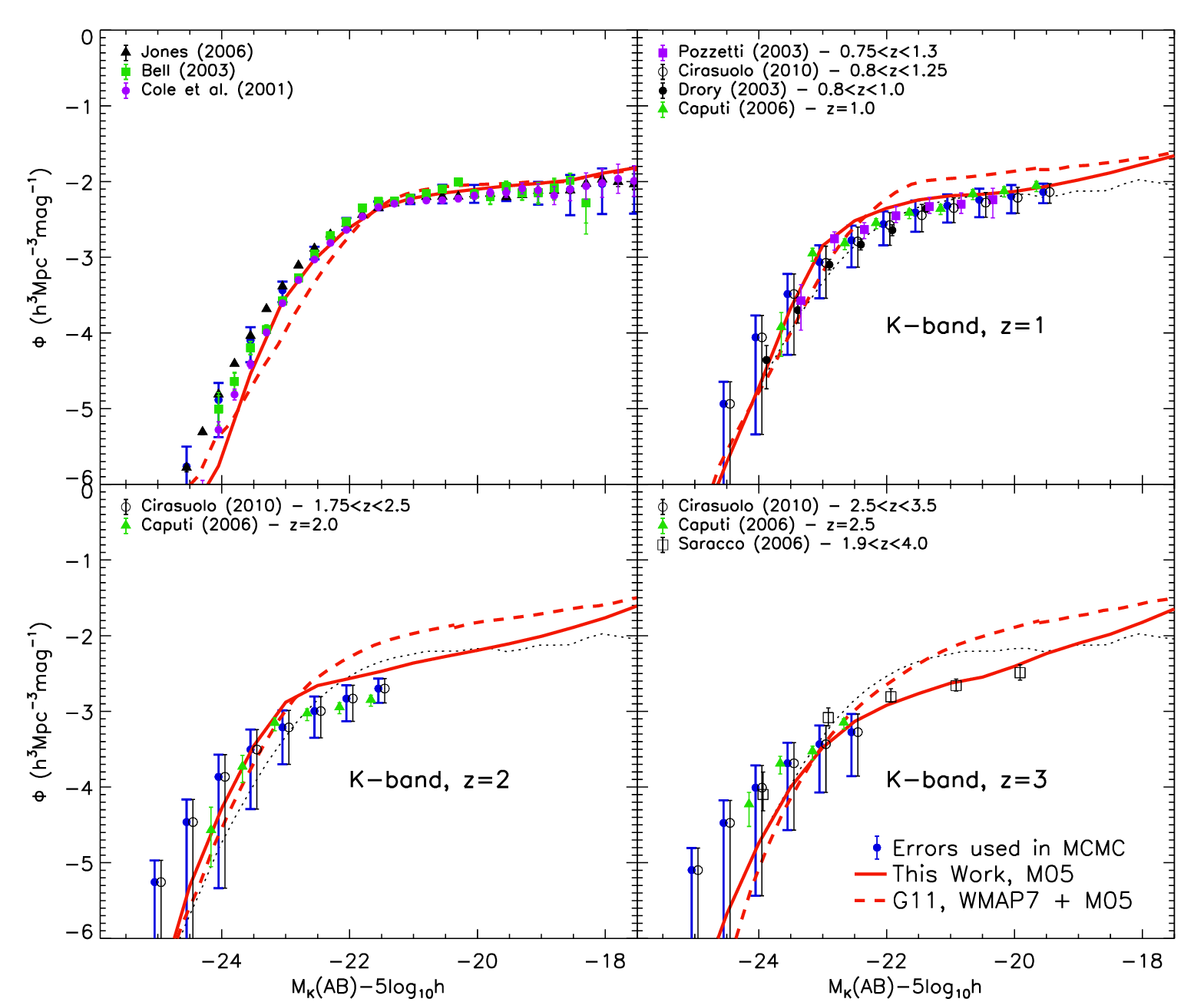}
\caption{Evolution of the rest-frame $K$-band luminosity function from
  $z=3$ to $z=0$ as in Fig.~\ref{fig:kband}, except that data points
  for the individual surveys are shown. These are three different
  2MASS-based determinations at $z=0$ \citep{Cole2001, Bell2003,
    Jones2006}, and determinations at higher redshift from MUNICS
  \citep{Drory2003}, the K20 survey \citep{Pozzetti2003}, the HDF-S
  \citep{Saracco2006}, GOODS-CDFS \citep{Caputi2006}, and the
  UKIDSS-UDF \citep{Cirasuolo2010}. The $z=0$ results of
  \citet{Jones2006} are repeated at all redshifts as a black dotted
  line.}
\label{fig:kband_obs}
\end{figure*}

In this Appendix we discuss in more detail our method for combining
individual datasets to construct observational constraints with
realistic error bars for each of the population properties used in the
MCMC sampling presented in section \ref{sec:results}. This is a
crucial aspect of the MCMC analysis as the allowed regions in
parameter space and the goodness of a fit to multiple properties
depend strongly on the observational uncertainties. While errors based
on counting statistics and cosmic variance are given in most
observational studies it is much harder to get a realistic assessment
of systematic uncertainties. These can show up as apparent
inconsistencies between different determinations of the same
population property, and failing to account for them can jeopardise a
meaningful comparison with theoretical predictions.

Here we follow \citet{Henriques2009} and \citet{Henriques2010}, using
multiple determinations of each observational property and taking the
scatter among them to indicate likely systematic uncertainties. For
each observational property, at each redshift, we re-bin the
individual estimates into a fixed set of broad bins. In each bin we
then discount any determination with very large error bars and take a
straight average of the rest as the constraint and their maximum and
minimum values as its $\pm 1\sigma$ uncertainty.  By not weighting the
averages we attempt to account for the fact that systematic errors can
affect large and small surveys in similar ways.  This can be seen, for
example, in the work of \citet{Marchesini2009} and
\citet{Marchesini2010} for the stellar mass functions and
\citet{Cirasuolo2010} for the $K$-band luminosity function.  Counting
errors are swamped in their overall error budget by combined
uncertainties from SED fitting assumptions, photometric redshift
errors, photometric errors, extrapolations from the observed
photometry and cosmic variance.  

Finally we harmonise the sizes of error bars to avoid dramatic changes
between bins, for example, where there is a change in the number of
surveys included.  Typically, we re-size the error bar to the value of
the next consecutive bin towards the knee of the function. More
surveys are normally available towards the knee and their counting
errors are usually smaller there, making systematics estimates more
robust in this region. Clearly our procedure is an art rather than a
quantitative science, but we believe it gives realistic estimates of
the overall level of uncertainty given the information available.

\begin{figure*}
\centering
\includegraphics[scale=0.5]{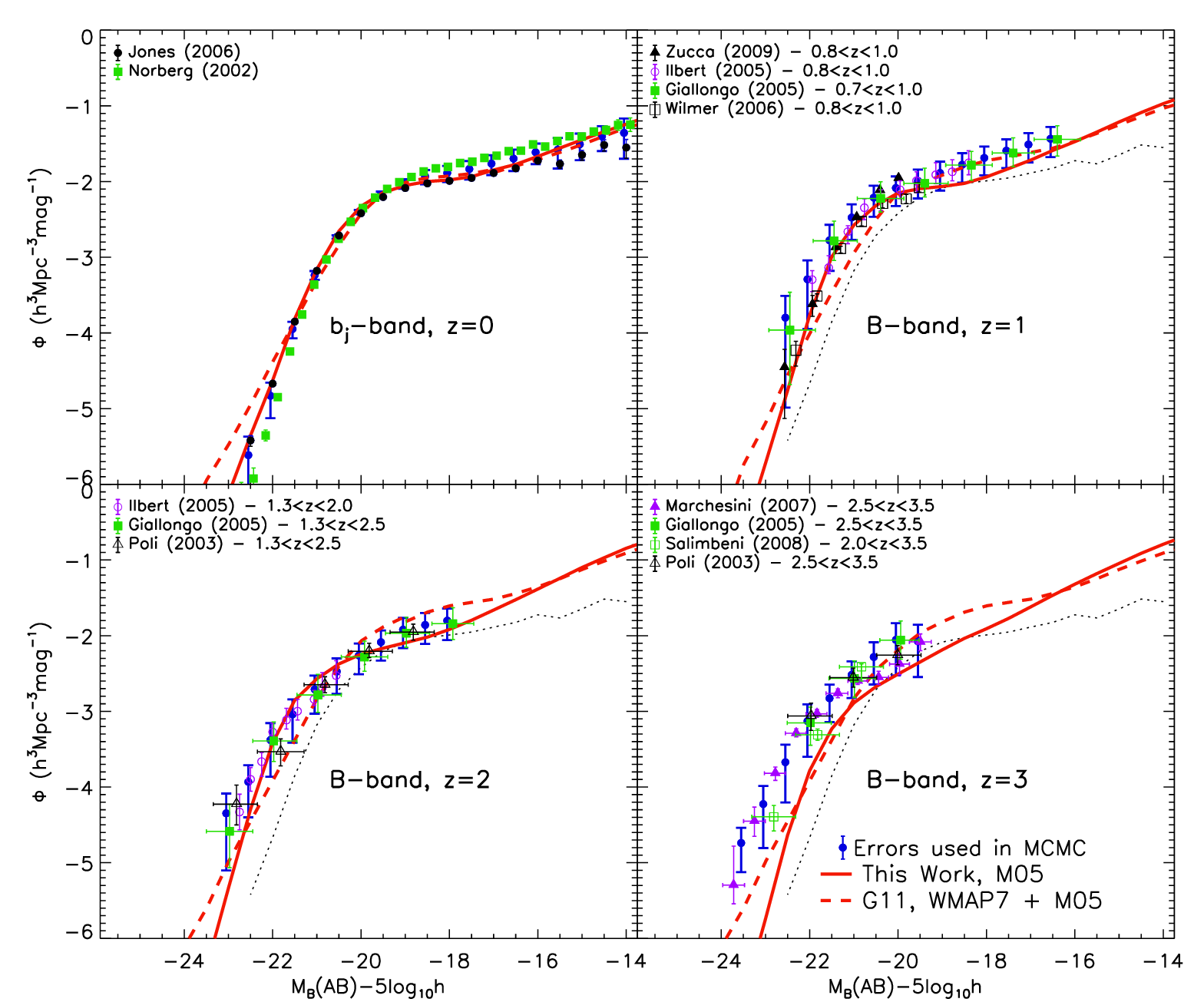}
\caption{Evolution of the rest-frame $B$-band luminosity function from
  $z=3$ to $z=0$ as in Fig.~\ref{fig:bband}, except that data points
  for the individual surveys are shown.  At $z=0$, these are the
  2DFGRS \citep{Norberg2002} and the 6DFGRS \citep{Jones2006}. At
  higher redshift, they are for HDF-S \citep{Poli2003}, HDF-N
  \citep{Giallongo2005}, VVDS \citep{Ilbert2005}, DEEP2
  \citep{Willmer2006}, GOODS-MUSYC plus FIRES \citep{Marchesini2007},
  GOODS-MUSYC \citep{Salimbeni2008}, and zCOSMOS \citep{Zucca2009}.
  The $z=0$ results of \citet{Jones2006} are repeated at all redshifts as a
  black dotted line.}
\label{fig:bband_obs}
\end{figure*}

Our adopted constraints are shown together with the individual
datasets on which they are based in Figure~\ref{fig:smf_obs} for
the stellar mass functions, in Figure~\ref{fig:kband_obs} for the
$K$-band luminosity functions, and in Figure~\ref{fig:bband_obs}
for the $B$-band luminosity functions. The constraints are shown as
blue dots with error bars while other data points represent
observational estimates from the individual surveys. Theoretical
predictions are shown as red lines. For the stellar mass function, the
most massive bin at $z=0$ and the most massive and least massive bins
at $z=1$ have been re-sized according to the method described in the
last paragraph. The same was done for the faintest bin in the $z=3$
$B$-band luminosity function.

We emphasise that despite our attempt to estimate uncertainties from
the data, our method still involves arbitrary judgements about the
quality of data and the way in which different surveys should be
combined. As a result, formal levels of agreement between theory and
observation should be treated with caution.

\label{lastpage}

\end{document}